\documentclass[twocolumn,trackchanges,twocolappendix]{aastex701}

\newcommand{\msun}{\mathrm{M}_\odot}

\defcitealias{2026ApJ..1000...33Q}{Q26}

\received{Oct. 4, 2026}
\revised{Nov. 12, 2026}
\accepted{Nov. 18, 2026}

\usepackage{graphicx}
\usepackage[T1]{fontenc}
\usepackage{aecompl}
\usepackage[]{amsmath}

\usepackage{color}

\usepackage{xspace}

\usepackage{soul}
\begin{document}

\title{The Binary Fraction of Milky Way Field Stars in DESI: Dependence on Chemical Abundances}

\author[0000-0002-4900-2088]{Tian Qiu}
\email{qtskyxt@sjtu.edu.cn}
\thanks{qtskyxt@sjtu.edu.cn}
\affiliation{Department of Astronomy, School of Physics and Astronomy, and Key Laboratory for Particle Astrophysics and Cosmology (MOE)/Shanghai Key Laboratory for Particle Physics and Cosmology, Shanghai Jiao Tong University, Shanghai 200240, People's Republic of China}
\affiliation{State Key Laboratory of Dark Matter Physics, School of Physics and Astronomy, Shanghai Jiao Tong University, Shanghai 200240, China}
\author[0000-0002-5762-7571]{Wenting Wang}
\email{wenting.wang@sjtu.edu.cn}
\thanks{wenting.wang@sjtu.edu.cn}
\affiliation{Department of Astronomy, School of Physics and Astronomy, and Key Laboratory for Particle Astrophysics and Cosmology (MOE)/Shanghai Key Laboratory for Particle Physics and Cosmology, Shanghai Jiao Tong University, Shanghai 200240, People's Republic of China}
\affiliation{State Key Laboratory of Dark Matter Physics, School of Physics and Astronomy, Shanghai Jiao Tong University, Shanghai 200240, China}
\author[0000-0003-2644-135X]{Sergey E. Koposov}
\email{Sergey.Koposov@ed.ac.uk}
\thanks{Sergey.Koposov@ed.ac.uk}
\affiliation{Institute for Astronomy, University of Edinburgh, Royal Observatory, Blackford Hill, Edinburgh EH9 3HJ, UK}
\affiliation{Institute of Astronomy, University of Cambridge, Madingley Road, Cambridge CB3 0HA, UK}
\author[0000-0002-6469-8263]{Songting Li}
\email{songting.li@sjtu.edu.cn}
\affiliation{Department of Astronomy, School of Physics and Astronomy, and Key Laboratory for Particle Astrophysics and Cosmology (MOE)/Shanghai Key Laboratory for Particle Physics and Cosmology, Shanghai Jiao Tong University, Shanghai 200240, People's Republic of China}
\author[0000-0002-7393-3595]{Nathan Sandford}
\email{nathan.sandford@utoronto.ca}
\affiliation{Department of Astronomy \& Astrophysics, University of Toronto, Toronto, ON M5S 3H4, Canada}
\author[0000-0002-9110-6163]{Ting S. Li}
\email{ting.li@astro.utoronto.ca}
\affiliation{Department of Astronomy \& Astrophysics, University of Toronto, Toronto, ON M5S 3H4, Canada}
\author[0000-0002-5758-150X]{Joan Najita}
\email{joan.najita@noirlab.edu}
\affiliation{NSF NOIRLab, 950 N. Cherry Ave., Tucson, AZ 85719, USA}
\author{Jessica Nicole Aguilar}
\email{jaguilar@lbl.gov}
\affiliation{Lawrence Berkeley National Laboratory, 1 Cyclotron Road, Berkeley, CA 94720, USA}
\author[0000-0001-6098-7247]{Steven Ahlen}
\email{ahlen@bu.edu}
\affiliation{Department of Physics, Boston University, 590 Commonwealth Avenue, Boston, MA 02215 USA}
\author[0000-0001-9712-0006]{Davide Bianchi}
\email{davide.bianchi1@unimi.it}
\affiliation{Dipartimento di Fisica ``Aldo Pontremoli'', Universit\`a degli Studi di Milano, Via Celoria 16, I-20133 Milano, Italy}
\affiliation{INAF-Osservatorio Astronomico di Brera, Via Brera 28, 20122 Milano, Italy}
\author{David Brooks}
\email{david.brooks@ucl.ac.uk}
\affiliation{Department of Physics \& Astronomy, University College London, Gower Street, London, WC1E 6BT, UK}
\author[0000-0003-3044-5150]{Aurelio Carnero Rosell}
\email{acarnero@iac.es}
\affiliation{Instituto de Astrof\'{\i}sica de Canarias, C/ V\'{\i}a L\'{a}ctea, s/n, E-38205 La Laguna, Tenerife, Spain}
\affiliation{Departamento de Astrof\'{\i}sica, Universidad de La Laguna (ULL), E-38206, La Laguna, Tenerife, Spain}
\author{Todd Claybaugh}
\email{tmclaybaugh@lbl.gov}
\affiliation{Lawrence Berkeley National Laboratory, 1 Cyclotron Road, Berkeley, CA 94720, USA}
\author[0000-0002-2169-0595]{Andrei Cuceu}
\email{acuceu@lbl.gov}
\affiliation{Lawrence Berkeley National Laboratory, 1 Cyclotron Road, Berkeley, CA 94720, USA}
\author[0000-0002-1769-1640]{Axel de la Macorra}
\email{macorra@fisica.unam.mx}
\affiliation{Instituto de F\'{\i}sica, Universidad Nacional Aut\'{o}noma de M\'{e}xico, Circuito de la Investigaci\'{o}n Cient\'{\i}fica, Ciudad Universitaria, Cd. de M\'{e}xico C.~P.~04510, M\'{e}xico}
\author[0000-0002-4928-4003]{Arjun Dey}
\email{arjun.dey@noirlab.edu}
\affiliation{NSF NOIRLab, 950 N. Cherry Ave., Tucson, AZ 85719, USA}
\author{Peter Doel}
\email{apd@star.ucl.ac.uk}
\affiliation{Department of Physics \& Astronomy, University College London, Gower Street, London, WC1E 6BT, UK}
\author[0000-0002-2890-3725]{Jaime E. Forero-Romero}
\email{je.forero@uniandes.edu.co}
\affiliation{Departamento de F\'isica, Universidad de los Andes, Cra. 1 No. 18A-10, Edificio Ip, CP 111711, Bogot\'a, Colombia}
\affiliation{Observatorio Astron\'omico, Universidad de los Andes, Cra. 1 No. 18A-10, Edificio H, CP 111711 Bogot\'a, Colombia}
\author[0000-0001-9632-0815]{Enrique Gaztañaga}
\email{gaztanaga@gmail.com}
\affiliation{Institut d'Estudis Espacials de Catalunya (IEEC), c/ Esteve Terradas 1, Edifici RDIT, Campus PMT-UPC, 08860 Castelldefels, Spain}
\affiliation{Institute of Cosmology and Gravitation, University of Portsmouth, Dennis Sciama Building, Portsmouth, PO1 3FX, UK}
\author[0000-0003-3142-233X]{Satya Gontcho A Gontcho}
\email{satya@virginia.edu}
\affiliation{Lawrence Berkeley National Laboratory, 1 Cyclotron Road, Berkeley, CA 94720, USA}
\affiliation{University of Virginia, Department of Astronomy, Charlottesville, VA 22904, USA}
\author{Gaston Gutierrez}
\email{gaston@fnal.gov}
\affiliation{Fermi National Accelerator Laboratory, PO Box 500, Batavia, IL 60510, USA}
\author[0000-0002-0000-2394]{Stephanie Juneau}
\email{stephanie.juneau@noirlab.edu}
\affiliation{NSF NOIRLab, 950 N. Cherry Ave., Tucson, AZ 85719, USA}
\author{Robert Kehoe}
\email{kehoe@physics.smu.edu}
\affiliation{Department of Physics, Southern Methodist University, 3215 Daniel Avenue, Dallas, TX 75275, USA} 
\author[0000-0003-1838-8528]{Martin Landriau}
\email{mlandriau@lbl.gov}
\affiliation{Lawrence Berkeley National Laboratory, 1 Cyclotron Road, Berkeley, CA 94720, USA}
\author[0000-0001-7178-8868]{Laurent Le Guillou}
\email{llg@lpnhe.in2p3.fr}
\affiliation{Sorbonne Universit\'{e}, CNRS/IN2P3, Laboratoire de Physique Nucl\'{e}aire et de Hautes Energies (LPNHE), FR-75005 Paris, France}
\author[0000-0002-1125-7384]{Aaron Meisner}
\email{aaron.meisner@noirlab.edu}
\affiliation{NSF NOIRLab, 950 N. Cherry Ave., Tucson, AZ 85719, USA}
\author{Ramon Miquel}
\email{rmiquel@ifae.es}
\affiliation{Institut de F\'{i}sica d’Altes Energies (IFAE), The Barcelona Institute of Science and Technology, Edifici Cn, Campus UAB, 08193, Bellaterra (Barcelona), Spain}
\affiliation{Instituci\'{o} Catalana de Recerca i Estudis Avan\c{c}ats, Passeig de Llu\'{\i}s Companys, 23, 08010 Barcelona, Spain}
\author[0000-0002-4637-2868]{Enrique Paillas}
\email{enrique.paillas@udp.cl}
\affiliation{Instituto de Estudios Astrof\'isicos, Facultad de Ingenier\'ia y Ciencias, Universidad Diego Portales, Av. Ej\'ercito Libertador 441, Santiago, Chile}
\author[0000-0002-0644-5727]{Will Percival}
\email{will.percival@uwaterloo.ca}
\affiliation{Department of Physics and Astronomy, University of Waterloo, 200 University Ave W, Waterloo, ON N2L 3G1, Canada}
\affiliation{Perimeter Institute for Theoretical Physics, 31 Caroline St. North, Waterloo, ON N2L 2Y5, Canada}
\affiliation{Waterloo Centre for Astrophysics, University of Waterloo, 200 University Ave W, Waterloo, ON N2L 3G1, Canada}
\author[0000-0001-7145-8674]{Francisco Prada}
\email{fprada@iaa.es}
\affiliation{Instituto de F\'{i}sica Te\'{o}rica (IFT) UAM/CSIC, Universidad Aut\'{o}noma de Madrid, Cantoblanco, E-28049, Madrid, Spain}
\affiliation{Instituto de Astrof\'{i}sica de Andaluc\'{i}a (CSIC), Glorieta de la Astronom\'{i}a, s/n, E-18008 Granada, Spain}
\affiliation{Instituto de Astrof\'{\i}sica de Canarias, C/ V\'{\i}a L\'{a}ctea, s/n, E-38205 La Laguna, Tenerife, Spain}
\author[0000-0001-6979-0125]{Ignasi P\'erez-R\`afols}
\email{ignasi.perez.rafols@upc.edu}
\affiliation{Departament de F\'isica, EEBE, Universitat Polit\`ecnica de Catalunya, c/Eduard Maristany 10, 08930 Barcelona, Spain}
\author[0000-0002-3500-6635]{Corentin Ravoux}
\email{corentin.ravoux@clermont.in2p3.fr}
\affiliation{Universit\'{e} Clermont-Auvergne, CNRS, LPCA, 63000 Clermont-Ferrand, France}
\author{Graziano Rossi}
\email{graziano@sejong.ac.kr}
\affiliation{Department of Physics and Astronomy, Sejong University, 209 Neungdong-ro, Gwangjin-gu, Seoul 05006, Republic of Korea}
\author[0000-0002-9646-8198]{Eusebio Sanchez}
\email{eusebio.sanchez@ciemat.es}
\affiliation{CIEMAT, Avenida Complutense 40, E-28040 Madrid, Spain}
\author[0000-0002-0408-5633]{Christoph Saulder}
\email{csaulder@mpe.mpg.de}
\affiliation{Max Planck Institute for Extraterrestrial Physics, Gie\ss enbachstra\ss e 1, 85748 Garching, Germany}
\author[0000-0002-3569-7421]{Edward Schlafly}
\email{eschlafly@stsci.edu}
\affiliation{Space Telescope Science Institute, 3700 San Martin Drive, Baltimore, MD 21218, USA}
\author{David Schlegel}
\email{djschlegel@lbl.gov}
\affiliation{Lawrence Berkeley National Laboratory, 1 Cyclotron Road, Berkeley, CA 94720, USA}
\author{Michael Schubnell}
\email{schubnel@umich.edu}
\affiliation{University of Michigan, 500 S. State Street, Ann Arbor, MI 48109, USA}
\affiliation{Department of Physics, University of Michigan, 450 Church Street, Ann Arbor, MI 48109, USA}
\author[0000-0002-3461-0320]{Joseph Harry Silber}
\email{jhsilber@lbl.gov}
\affiliation{Lawrence Berkeley National Laboratory, 1 Cyclotron Road, Berkeley, CA 94720, USA}
\author[0000-0002-2949-2155]{Małgorzata Siudek}
\email{msiudek@iac.es}
\affiliation{Instituto de Astrof\'{\i}sica de Canarias, C/ V\'{\i}a L\'{a}ctea, s/n, E-38205 La Laguna, Tenerife, Spain}
\author[0000-0003-1704-0781]{Gregory Tarl\'{e}}
\email{gtarle@umich.edu}
\affiliation{University of Michigan, 500 S. State Street, Ann Arbor, MI 48109, USA}
\author{Benjamin Alan Weaver}
\email{benjamin.weaver@noirlab.edu}
\affiliation{NSF NOIRLab, 950 N. Cherry Ave., Tucson, AZ 85719, USA}




\begin{abstract}

Binary populations encode information about the physical conditions of star formation and the subsequent evolution of stellar populations. Adopting a statistical method based on multi-epoch line-of-sight velocity variations, we constrain the binary fraction, $f_b$, of field main-sequence turn-off stars in the Milky Way using data from the second data release of the DESI Milky Way Survey. Our measurements are mainly sensitive to binaries with orbital periods shorter than approximately $10^4$ days. The inferred total binary fractions account statistically for less detectable systems through forward modeling under an assumed orbital-period distribution. $f_b$ is found to primarily increase with lower [Fe/H]. At $[\mathrm{Fe}/\mathrm{H}]\gtrsim -1.3$, $f_b$ shows a modest increase with increasing [Mg/Fe], whereas no clear trend is detected at lower [Fe/H]. The apparent variations of $f_b$ with angular momentum and Galactic height are largely explained by their underlying [Fe/H] distributions, while the remaining variations broadly follow differences in [Mg/Fe]. Within overlapping regions of the [Fe/H]--[Mg/Fe] plane, the thin disk, thick disk, stellar halo and GSE populations exhibit no significant differences in their binary fractions. Thus, our measurements indicate that the variation in $f_b$ is dominated by chemical abundance, which could be explained by the metallicity-dependent gas cooling efficiency and fragmentation. The weaker [Mg/Fe] dependence at fixed [Fe/H] might be associated with the formation conditions of stars, with higher [Mg/Fe] stars generally forming earlier under a more gas-rich and turbulent environment.

\end{abstract}




\section{Introduction} 
\label{sec:intro}

Binary stars are fundamental tracers of star formation, stellar evolution, and the accretion history of galaxies \citep{2001ApJ...556..265W,2013ARA&A..51..269D, 2018ApJ...854..147B}.
The frequency of binary systems encodes information about the star-forming region, while subsequent dynamical evolution can modify the surviving binary population over time \citep{1992PASP..104..981H,2011A&A...528A.144K,2020MNRAS.496.5176D}.
As a result, measurements of binary fraction across different stellar populations provide a useful probe of how stars form and evolve in different Galactic environments.

Studies in the solar neighborhood have shown that binary systems are common among Milky Way field stars \citep{1969JRASC..63..275H,1991A&A...248..485D,2010ApJS..190....1R}.
However, many studies are limited to relatively nearby volumes and therefore sample only a restricted range of stellar populations. 
Extending binary studies beyond the local solar neighborhood is essential for understanding whether binary fraction is universal or varies with stellar age, metallicity, and Galactic environment.

Several observational and theoretical studies have suggested that binary fraction may depend on chemical abundance. 
In particular, metallicity can influence binary formation through competing thermal processes. Lower metallicity reduces metal-line and dust cooling, but also decreases dust opacity, allowing radiation to escape more efficiently from sufficiently dense gas \citep{2019MNRAS.484.2341B, 2019ApJ...875...61M,2020MNRAS.499.1607M}. Radiation-hydrodynamical simulations predict that lower-metallicity gas has lower opacity and can cool more efficiently during collapse, promoting fragmentation and increasing the formation efficiency of close binaries, particularly at high densities where dust opacity becomes important.

\citet{2019MNRAS.490..550L} found that the binary fraction and mass-ratio distribution of solar-type field stars are strongly correlated, and interpreted this as evidence for early dynamical processing in embedded clusters, where low-binding-energy binaries with smaller masses, smaller mass ratios, and wider separations are preferentially disrupted. This highlights that present-day binary fractions reflect both the initial binary population and its subsequent evolution.
The formation of multiple-star systems is understood to involve a combination of turbulent fragmentation of molecular cloud cores, disk fragmentation, and subsequent dynamical evolution, with the relative importance of these processes depending on the physical conditions of the natal environment and the orbital separation of the system \citep{2023ASPC..534..275O}. Consequently, different binary populations, such as close and wide binaries, may exhibit distinct dependencies on metallicity and Galactic environment.
However, the behavior of binary fraction across the Milky Way, especially in chemically distinct populations spanning the disk and halo, remains insufficiently constrained.

Binary populations in the Galactic field have been studied for decades, beginning with nearby volume-limited samples in the solar neighborhood. Early astrometric and spectroscopic surveys established that multiplicity is common among nearby field stars \citep{1969JRASC..63..275H, 1991A&AS...88..281D}.
The benchmark studies of \citet{1991A&AS...88..281D, 1991A&A...248..485D} characterized the multiplicity of solar-type stars within the solar neighborhood and provided the widely used log-normal orbital-period distribution. 
This picture was later updated by \citet{2010ApJS..190....1R}, who presented a more complete census of nearby solar-type stars and found an overall multiplicity fraction of about 46\%, together with constraints on the period, and mass-ratio distributions.

With the advent of large spectroscopic surveys, binary studies moved beyond the immediate solar neighborhood and began to probe broader Galactic field populations. Using SDSS photometry, \citet{2015ApJ...799..135Y} developed a stellar-locus outlier method to statistically infer the binary fraction of main-sequence stars, finding an average binary fraction of about 40\% for field FGK stars. Multi-epoch radial-velocity surveys then enabled direct statistical constraints on unresolved binaries over much larger samples. Based on LAMOST DR4, \citet{2018RAA....18...52T,2020ApJS..249...22T} measured binary fractions for late-type stars using radial-velocity variability and found that the binary fraction increases with effective temperature and decreases with metallicity. They also reported that $\alpha$-enhanced populations tend to have higher binary fractions than low-$\alpha$ populations.  \citet{2022A&A...667A..44G} investigated the binary fraction of early-type stars and found that the binary fraction increases with effective temperature and metallicity with LAMOST DR8. APOGEE multi-epoch radial velocities provided a complementary view. \citet{2018ApJ...854..147B} showed that the distribution of maximum radial-velocity shifts depends strongly on $\log g$, and that metal-poor stars have multiplicity fractions about 2--3 times higher than metal-rich stars. This metallicity dependence was further synthesized by \citet{2019ApJ...875...61M}, who argued that the close binary fraction of solar-type stars is strongly anti-correlated with metallicity, decreasing by a factor of several from metal-poor halo-like populations to metal-rich disk populations. More recently, \citet{2024MNRAS.535..949B} investigated close-binary fractions in accreted and in-situ halo populations using Gaia DR3 radial-velocity variability. They found a moderately higher binary fraction among accreted stars, although the difference was not statistically significant, and showed that much of the variation with orbital energy and angular momentum could be associated with differences in metallicity. On the other hand, \citet{2025AJ....170..171L} used the DESI \textsc{SpecDis} distance catalog to identify unresolved equal-mass binary candidates and examined their occurrence across different chemical abundances. Interestingly, they found that the unresolved equal-mass binary had an opposite trend, showing an increasing binary fraction with increasing [Fe/H].

Resolved wide binaries provide a complementary probe of multiplicity at much larger separations. The {\it Gaia} era has made it possible to construct large samples of comoving wide binaries across the disk and halo. \citet{2019MNRAS.482L.139E} found that the metallicity dependence of the wide-binary fraction varies with projected separation, exhibiting a progressively stronger anticorrelation with [Fe/H] at smaller separations ($s\lesssim250$ au), while showing little dependence at larger separations. They also found that metal-poor binaries preferentially occupy smaller separations than their metal-rich counterparts, suggesting that the binary separation distribution itself depends on metallicity. \citet{2021MNRAS.501.4329H} revisited the metallicity dependence of field wide binaries using {\it Gaia} and LAMOST, showing that the wide-binary fraction has a non-monotonic dependence on metallicity in the disk. Extending this approach to the H3 survey, \citet{2022MNRAS.513..754H} compared thick-disk and halo wide binaries and found that, at fixed [Fe/H], the thick disk and halo have similar wide-binary fractions; their results also suggest that the wide-binary fraction is mainly linked to iron abundance rather than disk or halo origin or $\alpha$ abundance.

Despite this progress, several important questions remain. Nearby solar-neighborhood surveys provide the most precise binary statistics but probe only a limited range of Galactic environments. Large spectroscopic and astrometric surveys have extended these studies to broader stellar populations and revealed correlations between binary fraction and chemical abundance. However, the different observational techniques are sensitive to different regions of binary parameter space, for example, close versus wide systems or equal-mass versus general binaries, and are therefore subject to distinct selection effects. Here we apply a common stellar selection, multi-epoch line-of-sight velocities (LOSV) analysis, and binary-population model across a large field-star sample to examine the dependence of binary fraction on [Fe/H], [Mg/Fe], Galactic position, and Galactic component within a single framework.

In this work, we take advantage of the multi-epoch LOSV from the Dark Energy Spectroscopic Instrument (DESI) Milky Way Survey (MWS). Building on the statistical LOSV-variability framework applied in our previous DESI study of the Ursa Minor dwarf galaxy \citep{2026ApJ..1000...33Q}, we extend our method to a large sample of main-sequence turn-off (MSTO) field stars and measure the binary fraction across the [Fe/H]--[Mg/Fe] plane. This allows us to test whether the binary fraction is primarily associated with [Fe/H], [Mg/Fe], or Galactic position.

This paper is structured as follows: In Section~\ref{sec:data}, we describe the DESI observations, the construction of the MSTO sample, and the selection criteria adopted in this work.
Section~\ref{sec:method} presents the statistical framework used to infer the binary fraction, including the binary population model and the methodological updates introduced for the Galactic field sample. 
In Section~\ref{sec:results}, we present the inferred binary fractions across chemical abundance space and investigate their dependence on Galactic components. We discuss the implications of these results for binary formation and evolution in the Milky Way in Section~\ref{sec:discussion}. Finally, we summarize our main conclusions in Section~\ref{sec:conclusion}.

\section{Data}
\label{sec:data}

In this section, we describe the observational data utilized in this work.

\subsection{DESI observations}

DESI is mounted on the Mayall 4-meter telescope at Kitt Peak National Observatory, Arizona, and is designed to conduct wide-field spectroscopic surveys over approximately 14,000 deg$^2$ \citep{2013arXiv1308.0847L,2016arXiv161100036D, 2016arXiv161100037D, 2022AJ....164..207D,2023AJ....166..259S}. 
It is a multi-object spectroscopic instrument capable of collecting up to 5,000 spectra per exposure, providing a continuous wavelength coverage from 3,600 to 9,824~\AA~ with a full width at half maximum resolution of about 1.8~\AA~ \citep{2023AJ....165....9S,2023AJ....165..144G,2024AJ....168...95M,2024AJ....168..245P}.

Its primary cosmological survey aims to obtain spectra of more than 40 million galaxies and quasars to investigate the nature of dark energy 
\citep{2023AJ....165...50M,2023ApJ...943...68L,2023AJ....165..124A,2024AJ....167...62D}. 
The first Data Release \citep[DR1; ][]{2026AJ....171..285D} and the upcoming second Data Release (DR2) have already enabled, and will continue to deliver, a broad range of cosmological studies, including precise measurements of baryon acoustic oscillations (BAO) signals \citep{2025PhRvD.112h3515A} and increasingly stringent constraints on cosmological models \citep{2024arXiv241112022D}.

Alongside the cosmological surveys DESI conducted in
dark time under good observing conditions, the DESI Milky Way Survey \citep[MWS;][]{2023ApJ...947...37C} takes advantage of observing conditions suboptimal for galaxy redshift measurements, such as poor seeing and bright time with significant moonlight contamination. 
The primary DESI MWS program focuses on stars at Galactic latitudes $|b| > 20^\circ$, reaching up to a magnitude limit of $r = 19$. 
To date, the stellar value-added catalogs have been published based on the DESI EDR \citep{2024MNRAS.533.1012K} and DESI DR1 \citep{2026OJAp....955260K}, and DR2 will be made public soon. DR2 includes observations of more than 12 million stars selected from the DESI Legacy Imaging Survey \citep{2019AJ....157..168D}, and a significant number of stars have repeated observations. 

The DESI observations are organized into multiple survey phases—Commissioning (\textsc{cmx}), Survey Validation (\textsc{sv1}, \textsc{sv2}, \textsc{sv3}), the \textsc{main} survey, and dedicated \textsc{special} programs. 
Within each phase, targets are assigned to different observing programs (\textsc{dark}, \textsc{bright}, and \textsc{backup}), which correspond to distinct target selection algorithms and are executed under different observing conditions \citep{2024AJ....168...58D,2026AJ....171..285D}.

In this work, we utilize multi-epoch stellar spectra collected during the first three years of DESI MWS observations, which will be publicly released as part of the second data release (DR2) scheduled for spring 2027. The data reduction follows the DR1 stellar catalog described in \citet{2026OJAp....955260K}. These spectra provide measurements of LOSVs ($v_{\rm los}$), stellar atmospheric parameters, and chemical abundances. The LOSVs and atmospheric parameters, including surface gravity ($\log g$) and effective temperature ($T_{\rm eff}$), are derived using the MWS \textsc{rvs} pipeline\footnote{\url{https://github.com/segasai/rvspecfit}} \citep{2019ascl.soft07013K}. Further details of the MWS \textsc{rvs} pipeline are described in \citet{2023ApJ...947...37C}.

The chemical abundances used in this work, including [Fe/H] and [Mg/Fe], are adopted from the MWS \textsc{rvj} pipeline. The \textsc{rvj} pipeline is an extended version of \textsc{rvs} that provides measurements of multiple individual elemental abundances, including Carbon.
It employs a JAX-based spectral-fitting framework and synthetic spectral templates generated with \textsc{Korg}\footnote{\url{https://github.com/ajwheeler/Korg.jl}} \citep{2023AJ....165...68W,2024AJ....167...83W}, enabling abundance measurements over a wider metallicity range, including very metal-poor stars.

It is important to note that the uncertainties of LOSV ($\sigma_{v_{\rm{los}}}$) derived from the pipeline have a systematic floor of around 1-2~km~s$^{-1}$,
which refers to an irreducible component of measurement uncertainty that persists even when statistical errors are minimized.
It typically arises from wavelength calibration errors, which depend on observing conditions, and telescope condition changes such as dependence on the temperature.  This floor is modeled as a constant added in quadrature to the formal uncertainty and represents a lower limit to the achievable precision.
As discussed in Section~4.3.2 of \citet{2026OJAp....955260K}, the systematic error floors are 1.0, 1.6, and 2.0~km~s$^{-1}$ for the MWS \textsc{bright}, \textsc{dark}, and \textsc{backup} programs \citep[see][for details about the MWS \textsc{backup} program]{2025arXiv250517230D}, respectively.

In our analysis of this paper, we focus on the \textsc{bright} program. We will additionally constrain the systematic error floor as a function of stellar color, to improve the precision of our results in Section~\ref{ssec:syserr}.

\subsection{Main Sequence Turn-Off Star Selection}
\label{sec:MSTO}

To robustly constrain the binary fraction in the Galactic field, it is essential to probe regions with high stellar density while extending the spatial coverage to the outer disk and even into the stellar halo. We therefore select main-sequence turn-off (MSTO) stars as our primary sample. MSTO stars offer two key advantages: they are sufficiently numerous to provide strong statistical power, and their fundamental stellar parameters (e.g., mass, surface gravity, etc) can be determined with relatively high precision without requiring accurate age estimates.

We adopt the \textsc{main-bright} program as the parent sample for this analysis to avoid complicated selection functions introduced by combining different programs. Among the MWS observing programs, \textsc{main-bright} contains the largest sample and provides extensive multi-epoch coverage. 
Importantly, within a given observing program, observations are conducted under broadly similar observing conditions, including sky brightness and seeing constraints, etc. 
For objects belonging exclusively to a single target class, the probability of obtaining a DESI spectrum is therefore primarily determined by the number of survey passes over a given sky region and local fiber-assignment competition, rather than by intrinsic stellar properties such as magnitude, color, or proper motion. 
This combination of homogeneous observing conditions and selections that are largely independent of stellar parameters makes \textsc{main-bright} particularly well suited for binary fraction studies, as it minimizes systematic biases in the detection of LOSV variability. The formal definitions of target selections are given in detail in \citet{2023AJ....165...50M} and \citet{2023ApJ...947...37C}.

The \textsc{main-bright} program  targets objects in the magnitude range $16<r<19$. Within this program, we select the  \textsc{main-blue} subsample by applying the corresponding bitmask, which includes a color selection of $g-r<0.7$. Unless otherwise specified, all magnitudes referenced in this work are measured in the DESI Legacy Imaging Surveys photometric system \citep{2019AJ....157..168D}. To construct our parent sample, we require sources to be primary targets ($\mathrm{PRIMARY} = \mathrm{True}$) and spectroscopically classified as stars ($\mathrm{RR\_SPECTYPE} = \mathrm{STAR}$) by the \textsc{Redrock} pipeline \citep[Bailey et al. in preparation, ][]{2024AJ....168..124A}. In addition, we retain only spectra with reliable LOSV measurements by imposing $\mathrm{RVS\_WARN}=0$ and $\sigma_{v_{\rm los}}<20$ km~s$^{-1}$, thereby excluding observations flagged as unreliable or with poor velocity measurements. Here, $\sigma_{\rm vlos}$ denotes the formal uncertainty of an individual LOSV measurement reported by the pipeline, rather than the standard deviation of the LOSV measurements across multiple epochs. Because our analysis relies on multi-epoch LOSV measurements to identify velocity variability, we further exclude stars whose repeated exposures were obtained exclusively within a single night. This requirement ensures a non-zero temporal baseline for LOSV variability detection. We further exclude sources identified as variable in the {\it Gaia} DR3 variability summary catalog \citep[gaiadr3.vari\_summary, ][]{2023A&A...674A..13E, 2023A&A...674A...1G}. This leaves us with 343,147 stars. 

\begin{figure}
\plotone{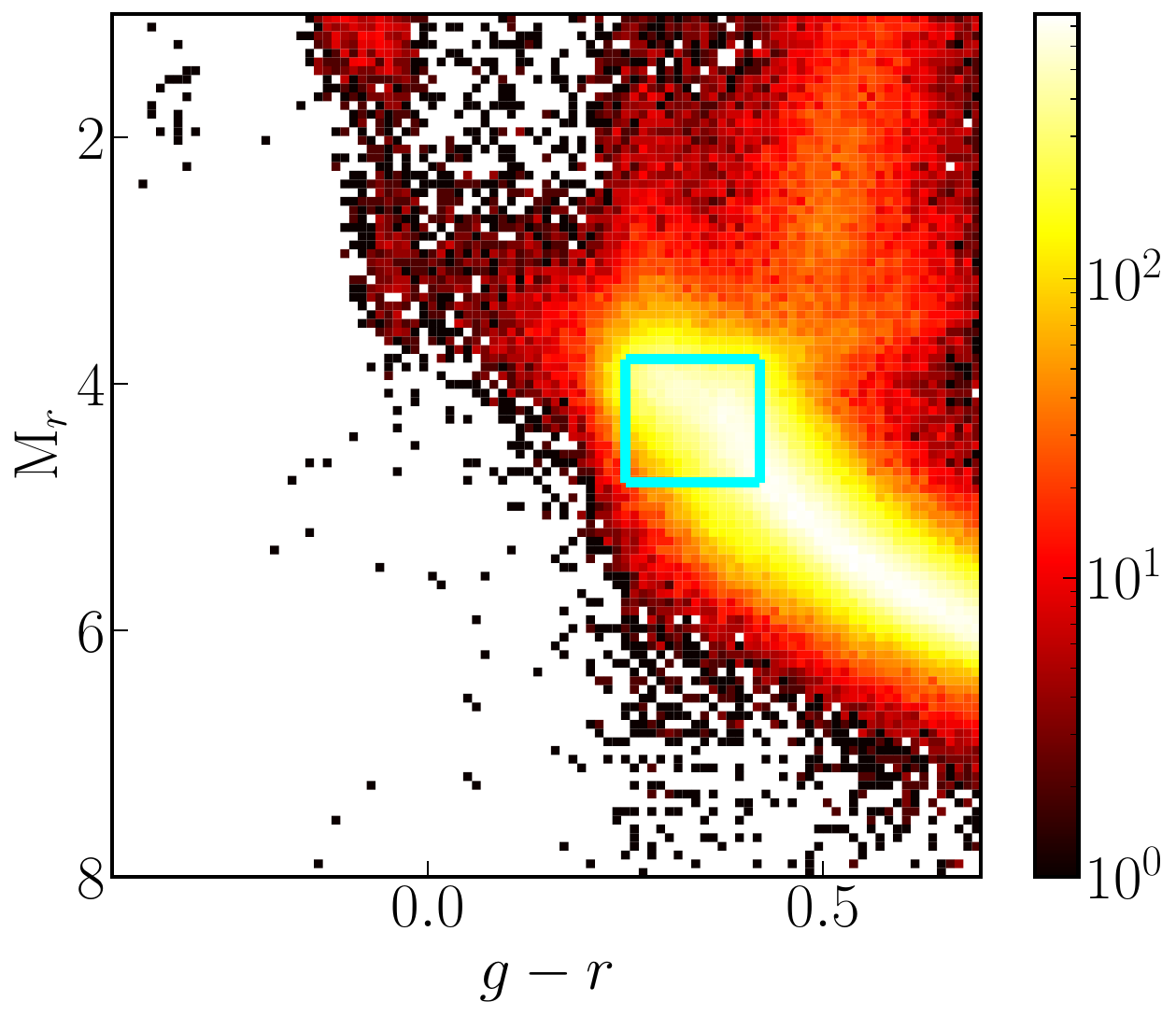}
\caption{Color–absolute magnitude diagram of the MWS \textsc{main-blue} sample after the initial quality cuts. The cyan box indicates the adopted MSTO selection region, defined by $3.8<\mathrm{M}_r<4.8$ and $0.25<g-r<0.42$.
\label{fig:cmd}}
\end{figure}

Figure~\ref{fig:cmd} presents the color–absolute magnitude diagram of the initial sample. Absolute magnitudes are computed using distances estimated from a neural network model developed by
\citet{2025AJ....170..171L}.
The selection criteria adopted for MSTO stars are given below,
\begin{equation}
\begin{aligned}
    3.8&<\mathrm{M}_r<4.8\\
    0.25&<g-r<0.42\\
    5000&<\mathrm{T}_\mathrm{eff}<6500 \ \mathrm{K}\\
    3.0&<\mathrm{log}\ g <4.5
\end{aligned}
\end{equation}
We also draw a cyan box in Figure~\ref{fig:cmd} to highlight our color and magnitude selection. The cuts on surface gravity and effective temperature are empirical and relatively loose, intended primarily to remove outliers with unreliable measurements.
This selection provides a sample of 81,339 MSTO stars.

\subsection{Distance Cut and Spatial Distribution}

\begin{figure}
\plotone{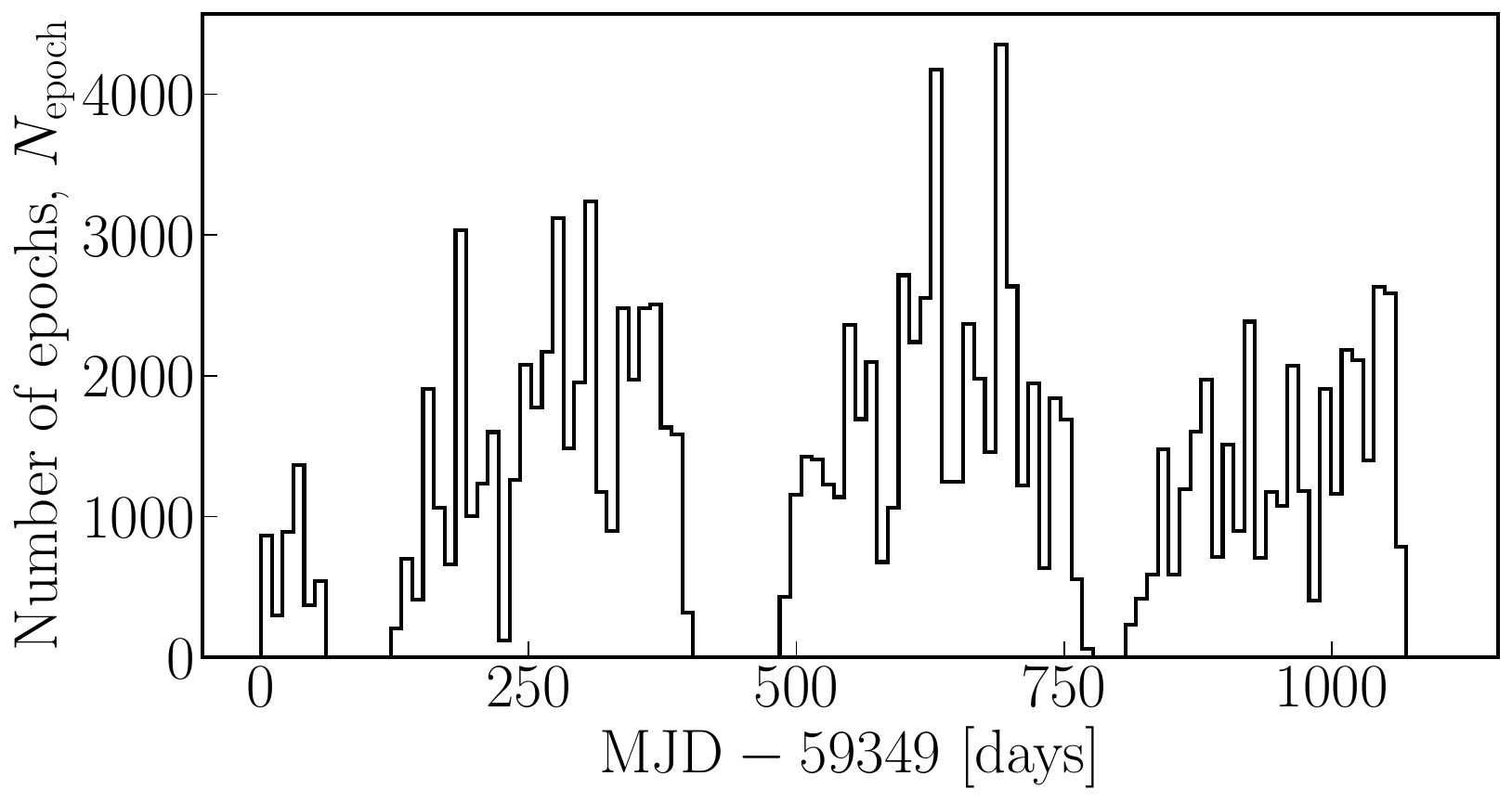}
\plotone{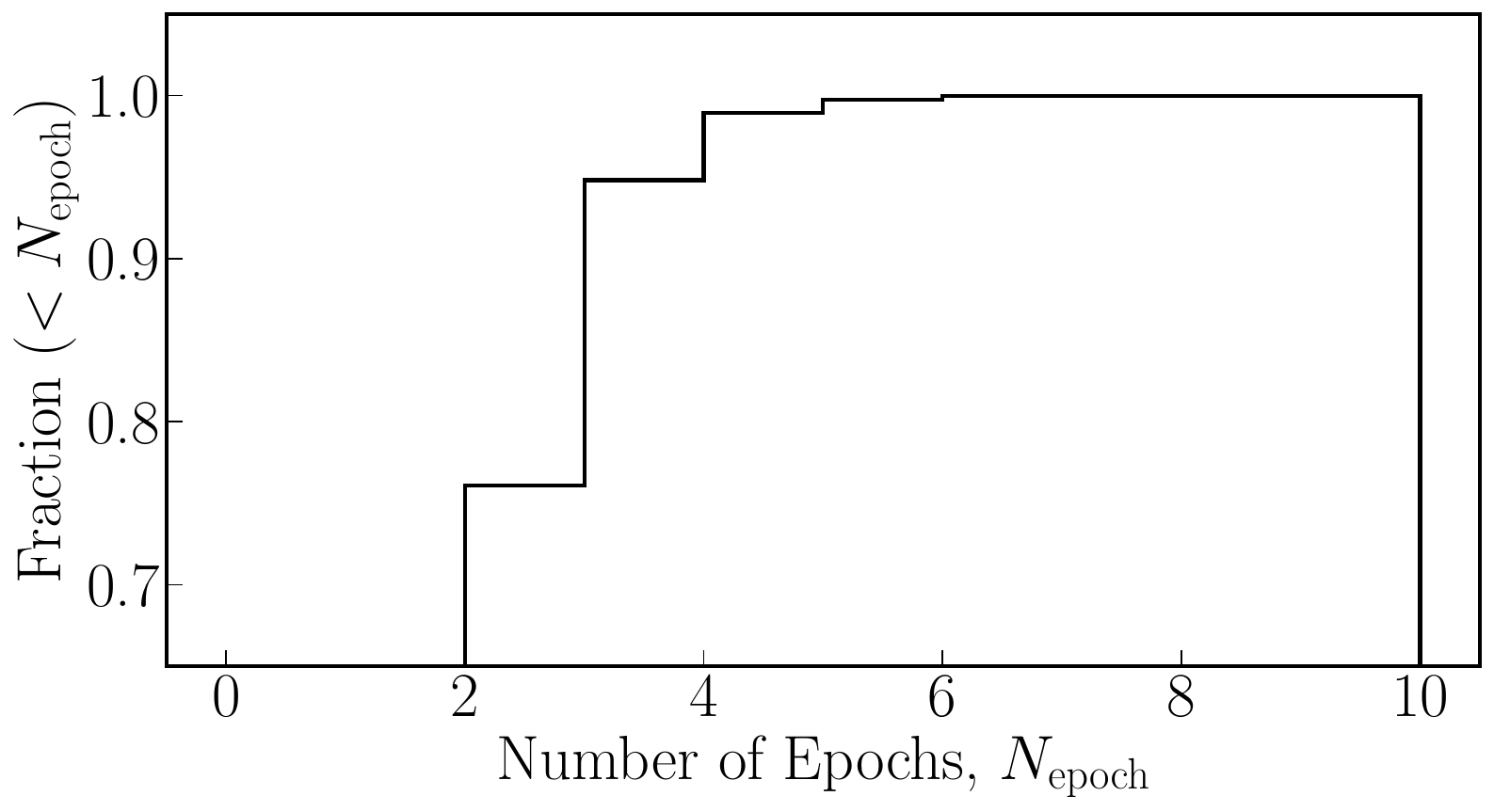}
\caption{The {\bf upper panel} shows the histogram of the Modified Julian Date (MJD) of all DESI observations included in this work. The observations span approximately three years and are distributed over multiple observing seasons, providing a wide range of baselines for measuring multi-epoch LOSV variations.
The {\bf lower panel} shows the cumulative histogram of the number of epochs for each star.
\label{fig:obshist}}
\end{figure}

\begin{figure}
\plotone{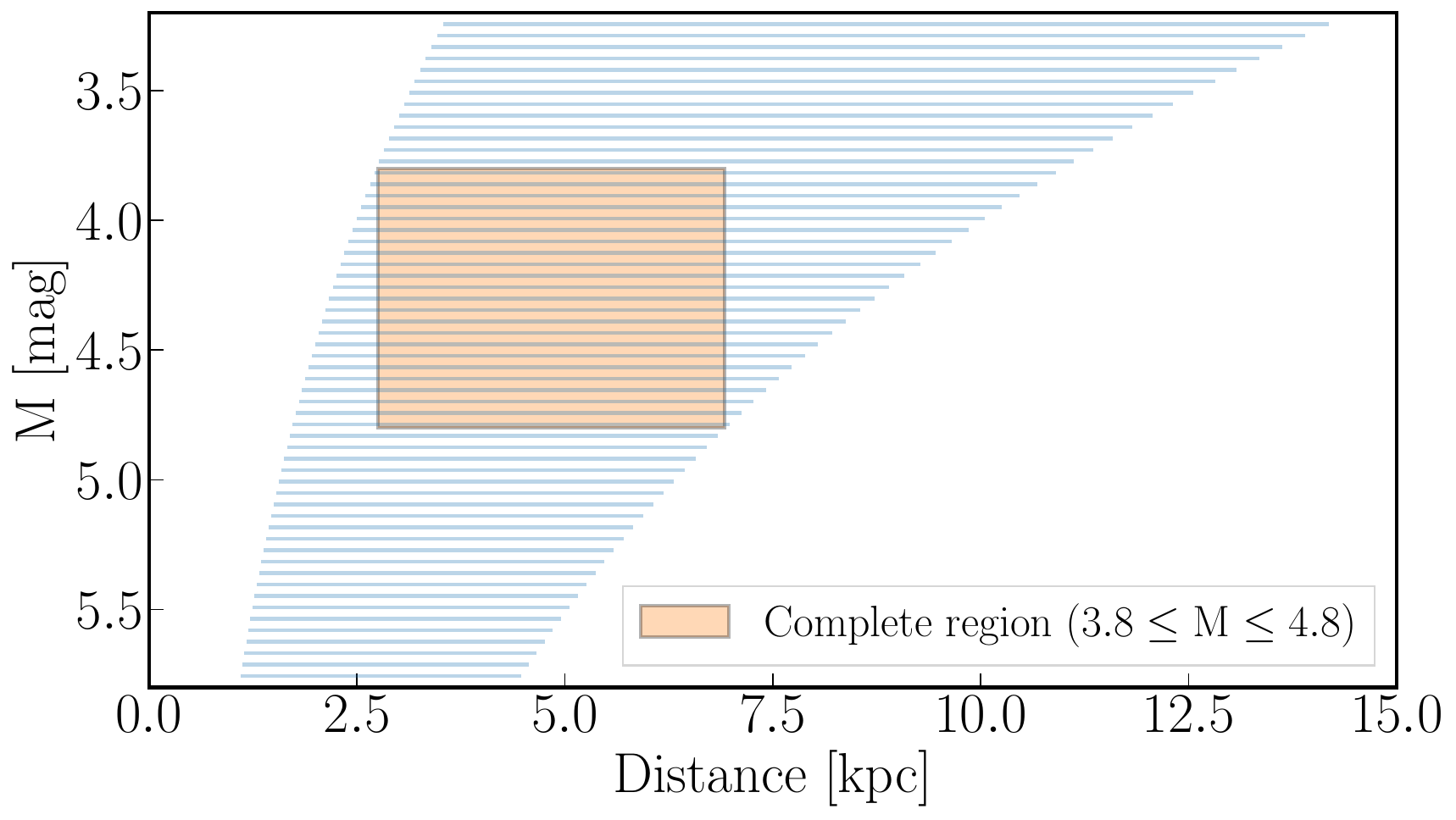}
\plotone{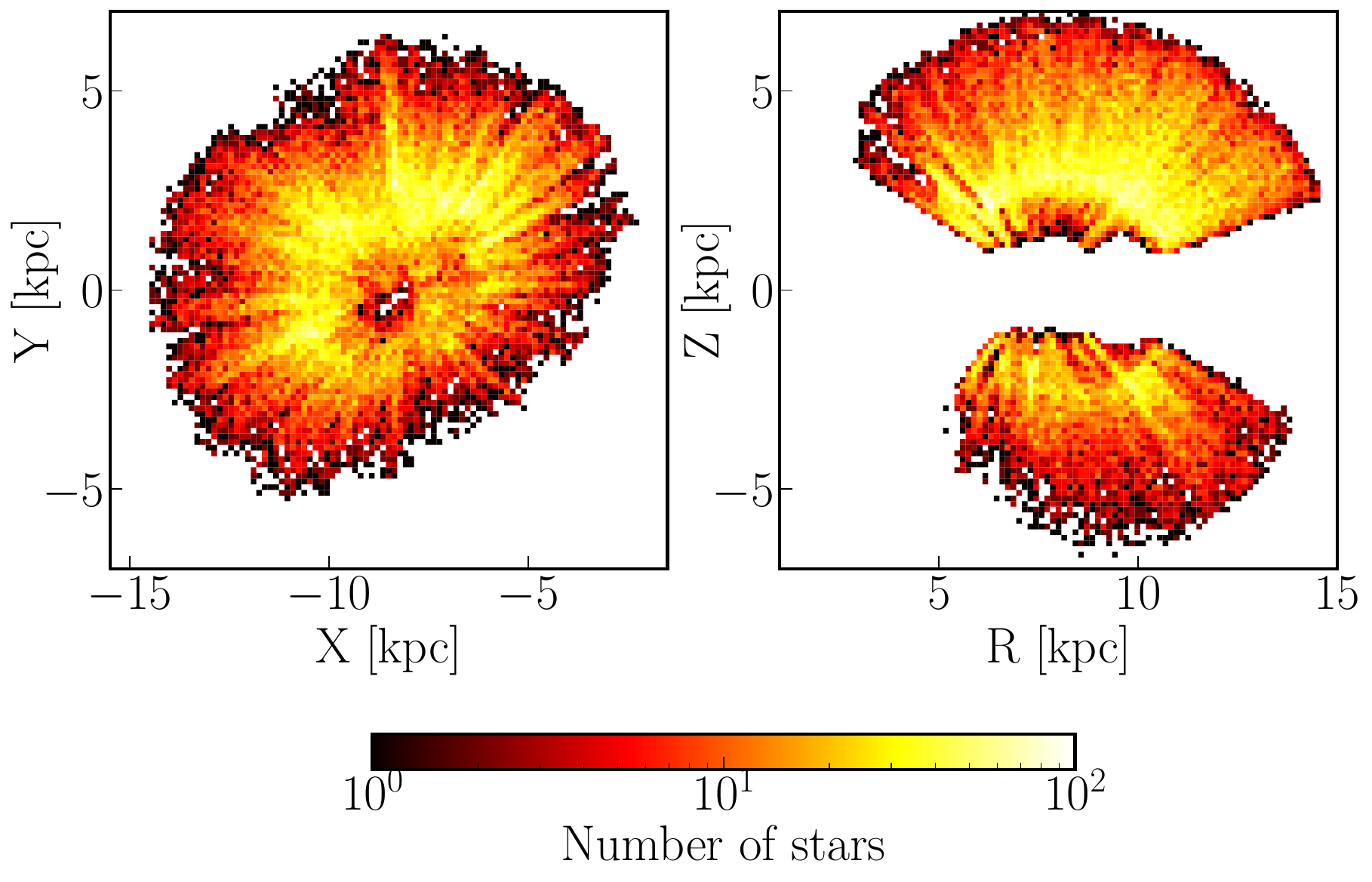}
\caption{\textbf{Upper panel:} Distance–absolute magnitude selection limits for the sample. The blue horizontal segments indicate, for each absolute magnitude M, the range of distances over which stars satisfy the apparent magnitude limits $16 \leq m \leq 19$. The shaded orange region highlights the distance interval within which stars with $3.8 \leq M \leq 4.8$ are fully observable across the entire magnitude range.\\
\textbf{Lower panels:} Spatial distribution of the MSTO sample after distance cut in Galactocentric coordinates.
{\bf Lower left:} distribution in the X–Y plane.
{\bf Lower right:} distribution in the R-Z plane, where $\mathrm{R} = \sqrt{\mathrm{X}^2 + \mathrm{Y}^2}$.
The color scale indicates the number of stars in each bin. The sample spans a heliocentric distance range of $\sim 2.75$–$6.92$ kpc, with coverage extending to several kiloparsecs above and below the Galactic plane.
\label{fig:distmag_spatial}}
\end{figure}

We further impose a distance cut of $2.75 < d < 6.92 $ kpc. This range is chosen such that, within the apparent magnitude limits of the \textsc{main-bright} program ($16 \leq r \leq 19$), the corresponding absolute magnitude coverage fully contains the MSTO selection region defined above, illustrated in the upper panel in Figure~\ref{fig:distmag_spatial}. As a result, this cut minimizes distance-dependent incompleteness and avoids introducing artificial color-dependent selection effects caused by the coupling between color and absolute magnitude. After applying this distance cut, the remaining volume limited sample is reduced to 57,647 stars with a total of 132,853 observations. The temporal distribution of the observations for the final sample is shown in Figure~\ref{fig:obshist}. The observations span approximately three years and are distributed over multiple observing seasons, yielding temporal baselines ranging from a few days to nearly three years for individual stars. Such a wide range of time baselines enables sensitivity to binary systems over a broad range of orbital periods through multi-epoch LOSV measurements.

The spatial distribution of this volume-limited MSTO sample is shown in the lower panels in Figure~\ref{fig:distmag_spatial}. In Galactocentric coordinates, the sample spans from 2.9 to 14.5 kiloparsecs in Galactocentric radius and extends about 6 kiloparsecs above and below the Galactic plane. Owing to the DESI MWS observing strategy, which targets fields at Galactic latitudes $|b|>20^\circ$ \citep{2023ApJ...947...37C}, relatively few stars are sampled near the mid-plane. Consequently, the sample is dominated by thick disk and halo populations, while the thin disk component is represented mainly by stars in the outer Galaxy, where the Galactic disk becomes increasingly flared \citep{2012ApJ...753..148B,2024MNRAS.531.1730T}.

\subsection{Chemical Abundance Selection and Binning}
\label{ssec:chemgrid}
\begin{figure}
\plotone{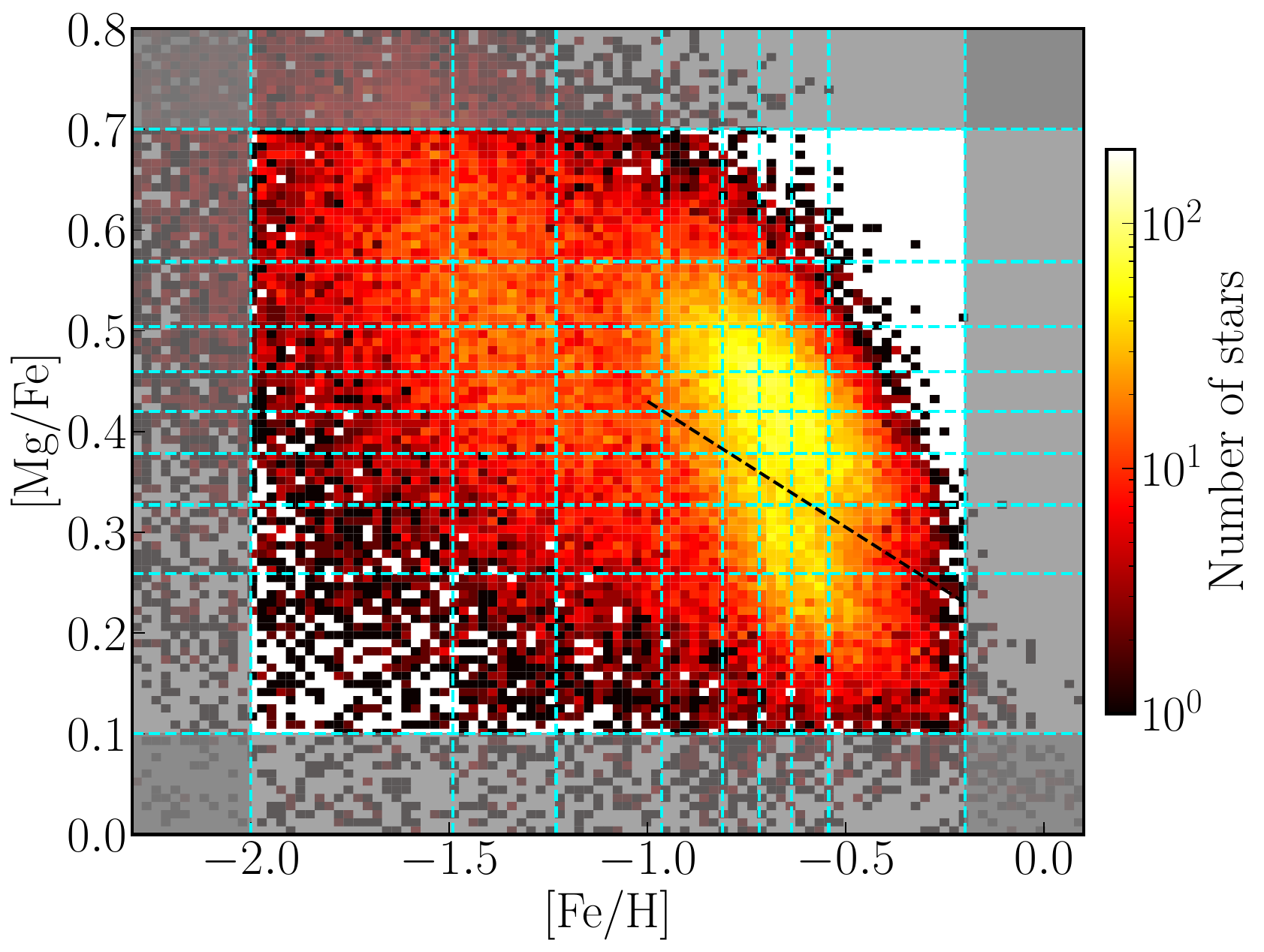}
\caption{Distribution of stars in the [Fe/H], [Mg/Fe] plane. The two-dimensional histogram shows the full MSTO sample and illustrates the underlying chemical abundance structure. We exclude the gray regions, where the stellar density is too low to provide meaningful constraints or enable robust comparisons across chemical-abundance bins.
Cyan dashed lines indicate the grids adopted for the binary fraction analysis.
The black dashed line indicates the chemical thin-/thick- disk separation which we adopt in Section~\ref{ssec:components}.
\label{fig:FeH_MgFe_grid}}
\end{figure}

The chemical abundance distribution of the selected sample is shown in Figure~\ref{fig:FeH_MgFe_grid}. Thick, thin disks and the halo components are clearly seen. This is similar to Figure~1 of \cite{2026ApJ...999..108L}, but is based on only the MSTO sample. To ensure reliable abundance measurements, we exclude stars with $[\mathrm{Fe}/\mathrm{H}] < -2.0$, where the inferred [Mg/Fe] values are less robust \citep{2026OJAp....955260K}. We further restrict the sample to $[\mathrm{Fe}/\mathrm{H}] < -0.2$ and $0.1 < [\mathrm{Mg}/\mathrm{Fe}] < 0.7$. Outside this region, the stellar density in the chemical abundance plane becomes too low to provide robust binary fraction constraints. Although larger bin size can be adopted in these sparse regions, the resulting measurements remain dominated by statistical uncertainties.

Stellar populations with different chemical abundances are associated with different Galactic components and formation histories. Therefore, within the resulting selection region, we subdivide the sample into bins in the [Fe/H]–[Mg/Fe] plane and infer the binary fraction for stars in each bin. Rather than adopting uniform abundance intervals, the bin boundaries are chosen so that each row and each column contains approximately equal numbers of stars according to the underlying stellar density. Because the occupied region in the abundance plane has an irregular shape, the resulting two-dimensional grid does not contain an equal number of stars in every cell. In particular, cells near the boundaries of the occupied region are generally larger in abundance extent and may contain either more or fewer stars than those in the central region, as illustrated in Figure~\ref{fig:FeH_MgFe_grid}. The resulting bins contain approximately 800 stars on average. To ensure sufficient statistical precision, we derive binary fractions only for bins containing at least 400 stars in Section~\ref{sec:results}.

\section{Methodology}
\label{sec:method}

We constrain the binary fraction using the variability of multi-epoch LOSVs, following the methodology developed in our previous study \citep[][hereafter \citetalias{2026ApJ..1000...33Q}]{2026ApJ..1000...33Q}. Rather than identifying individual binary systems, this approach treats the stellar sample statistically by characterizing the distribution of LOSV variability. We construct forward models of both binary and non-binary populations through Monte Carlo simulations, accounting for the observational sampling and LOSV measurement uncertainties. By comparing the observed variability distribution with the predictions of these models, we estimate the binary fraction through a likelihood-based framework.

For the binary population, we adopt the same orbital-parameter distributions as in \citetalias{2026ApJ..1000...33Q}, which are based on \citet{1991A&AS...88..281D} (hereafter DM91). Specifically, the orbital periods follow a log-normal distribution, $\frac{{\rm d}N}{{\rm d}\log(P/{\rm [day]})}\sim\mathcal{N}(4.8,2.3)$, where $\mu=4.8$ and $\sigma=2.3$ denote the mean and standard deviation in $\log(P/{\rm day})$, respectively. The mass ratios are drawn from a normal distribution $\frac{{\rm d}N}{{\rm d}q}\sim\mathcal{N}(0.23,0.42)$, where the mean is 0.23 and standard deviation is 0.42. The orbital inclinations are assumed to be isotropically distributed, corresponding to a probability density $\frac{{\rm d}N}{{\rm d}i}\propto\sin i$, while the eccentricity ($e$), argument of periastron ($\omega$), and mean anomaly ($M$) are each drawn from uniform distributions. We emphasize that adopting a universal orbital distribution allows a homogeneous relative comparison among populations, while possible abundance dependence of orbital properties remains an important uncertainty.

For each simulated realization, we generate LOSV measurements at the same observational epochs as the data and quantify the variability using a single summary statistic for each star. By comparing the observed variability distribution with those predicted by models with different assumed binary fractions, we infer the binary fraction through a likelihood-based framework combined with Markov Chain Monte Carlo sampling. Here we briefly summarize the main steps of the inference below.

We first quantify the LOSV difference between each pair of observations of a given star using the normalized statistic
\begin{equation}
    \alpha_{i,j} =\frac{|v_{{\rm los},i}-v_{{\rm los},j}|}{\sqrt{\sigma_{v_{\rm los},i}^{2}+\sigma_{v_{\rm los},j}^{2}}},
\end{equation}
where $v_{{\rm los},i,j}$ and $\sigma_{v_{\rm los},i,j}$ are the measured LOSV and its uncertainty at epoch $i,j$, respectively. The statistic $\alpha_{i,j}$ therefore represents the LOSV difference between two epochs scaled by the combined measurement uncertainty. For a non-variable star, the observed LOSV differences are expected to arise primarily from measurement uncertainties, whereas orbital motion can produce additional LOSV variations in binary systems.

To characterize the variability of each star using a single quantity, we combine all available pairs of observations for the same star into the statistic
\begin{equation}
\beta =1+\sqrt{\frac{2}{n(n-1)}\sum_{i<j}\alpha_{i,j}^{2}},
\end{equation}
where $n$ is the number of observational epochs. The normalization by the number of distinct epoch pairs ensures that stars with different numbers of observations can be characterized using the same statistic. The additive constant shifts the distribution to $\beta\geq1$, preventing values close to zero from accumulating at the lower boundary of the probability density distribution and facilitating a more stable representation of the distribution, particularly on a logarithmic scale. Larger values of $\beta$ indicate stronger LOSV variations relative to the measurement uncertainties, making binary systems with detectable orbital motion more likely to populate the high-$\beta$ tail.
We adopt this statistic because it combines the LOSV differences across all available epochs into a single uncertainty-normalized measure of variability. For equal measurement uncertainties, $\beta-1$ is equivalent to the sample standard deviation of the LOSVs divided by the individual measurement uncertainty, providing a straightforward interpretation of the observed variability relative to the expected measurement.

\begin{figure}
\plotone{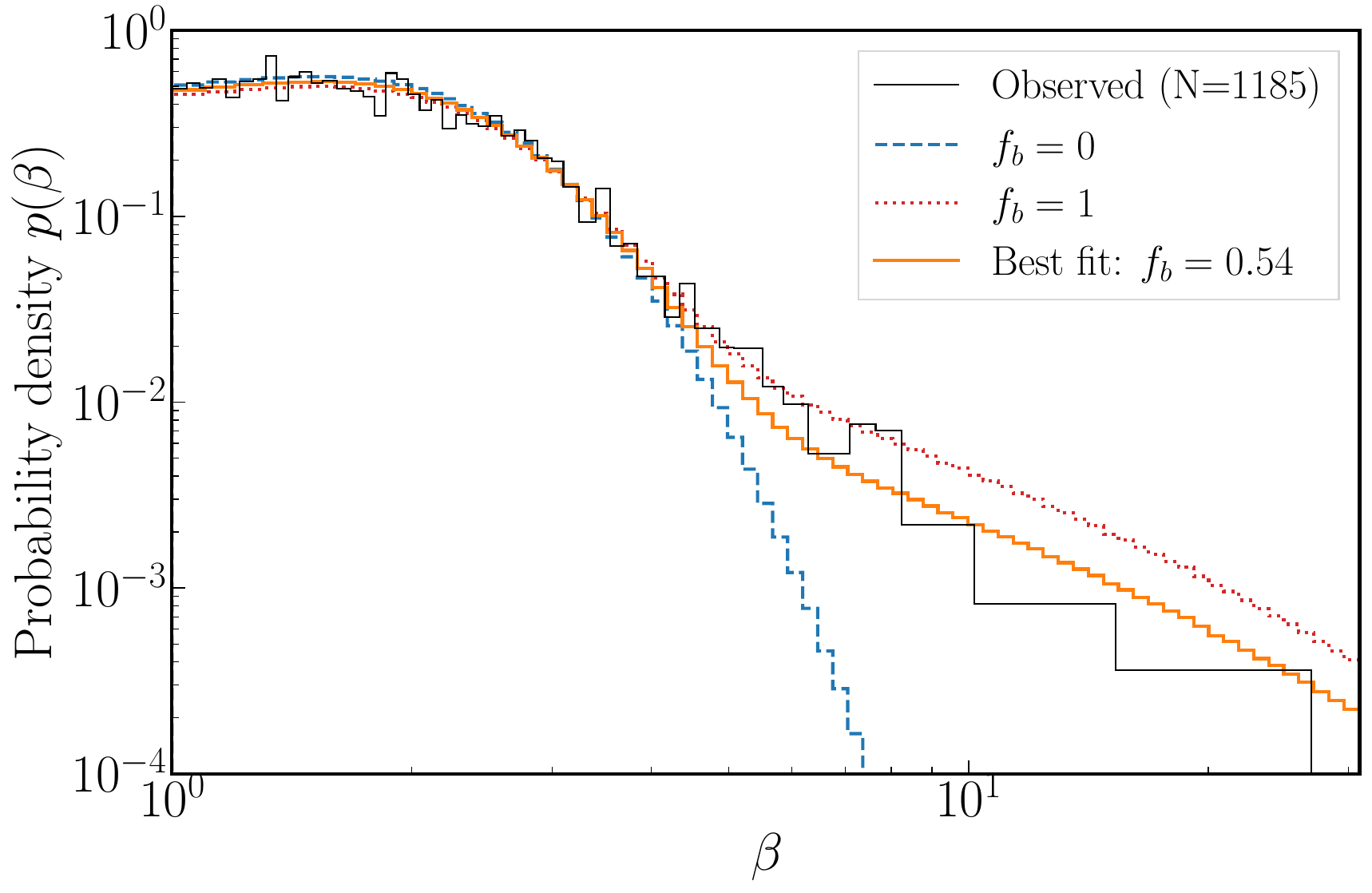}
\caption{Distribution of the LOSV variability $\beta$ for a representative subsample of 1,185 stars, selected from one of the chemical-abundance bins in Figure~\ref{fig:bf_grid}, spanning $-0.81 < [\mathrm{Fe/H}] < -0.72$ and $0.42 < [\mathrm{Mg/Fe}] < 0.46$. The black histogram shows the observed distribution with adaptive binning. The blue dashed and red dotted curves show the predicted probability densities for binary fractions $f_b=0$ and $f_b=1$, respectively. The orange solid curve represents the maximum-likelihood mixture model with $f_b=0.54$.
\label{fig:beta_dist}}
\end{figure}

We use Monte Carlo simulations to reproduce the expected LOSV variability distributions of both binary and non-binary populations, accounting for the orbital motions of binary systems, observational sampling, and measurement uncertainties. From these simulations, we construct the probability density distributions $P(\beta|f_b=1)$ and $P(\beta|f_b=0)$, corresponding to populations consisting entirely of binary and non-binary stars, respectively. As shown in Figure~\ref{fig:beta_dist}, the simulated non-binary population is concentrated at relatively low $\beta$ values, while the binary population exhibits a more extended high-$\beta$ tail due to orbital motion. By combining these two distributions with different relative contributions, we can reproduce the expected variability distribution for a population with any assumed binary fraction $f_b$, expressed as
\begin{equation}
P(\beta|f_b)=f_b P(\beta|f_b=1)+(1-f_b)P(\beta|f_b=0).
\end{equation}

Figure~\ref{fig:beta_dist} illustrates this procedure by comparing the observed $\beta$ distribution with the best-fitting mixture model. The observed distribution is well reproduced by combining the two simulated components with an appropriate binary fraction. By fitting the full $\beta$ distribution, our method statistically accounts for binary systems exhibiting weak LOSV variations, without requiring the identification of individual binary candidates based on a fixed variability threshold.

For a given binary fraction $f_b$, the likelihood of the full sample is calculated as
\begin{equation}
\mathcal{L}(f_b|\boldsymbol{\beta}_{\rm obs})=\prod_k P(\beta_{k,\rm obs}|f_b),
\end{equation}
where $\beta_{k,\rm obs}$ is the observed variability statistic of star $k$, and $P(\beta_{k,\rm obs}|f_b)$ is evaluated using the mixture distribution defined above. We then use Markov Chain Monte Carlo sampling to obtain the posterior distribution of $f_b$, from which the binary fraction and its uncertainty are determined.

The accuracy of this inference procedure was also tested through injection–recovery experiments in Section~4.4 of \citetalias{2026ApJ..1000...33Q}, in which mock samples with input binary fractions can be successfully recovered within their posterior uncertainties.

On the basis of \citetalias{2026ApJ..1000...33Q}, several aspects have been updated for the present analysis. In the following subsections, we describe the main changed aspects.

\subsection{Mass of Primary, $m_1$}

In \citetalias{2026ApJ..1000...33Q}, the primary mass, $m_1$, was not explicitly measured. Instead, we first drew an initial mass for each mock star from an assumed initial mass function (IMF), and then mapped it onto an isochrone to obtain the corresponding present-day stellar mass and other stellar parameters. This procedure was appropriate for UMi, whose member stars can be better approximated as a single stellar population.

In contrast to the UMi sample considered in \citetalias{2026ApJ..1000...33Q}, the Galactic field stars analyzed here span a broad range of chemical abundances, ages, and formation histories, making it inappropriate to approximate them as a single stellar population. We therefore adopt a different strategy and extend the \textsc{SpecDis} framework of \citet{2025AJ....170..171L} to estimate stellar masses in a data-driven way. In this extended version, \textsc{SpecDis2}, we first infer the stellar distances, and then combine these distances with spectroscopic parameters and photometric information to place each star on stellar isochrones and estimate its mass through interpolation. This approach is similar to the methods adopted by \citet{2023Natur.613..460L,2026ApJ...998L..33L}, in which stellar masses or mass functions are constrained by linking observed stellar properties to stellar evolution models.

Despite this diversity in stellar populations, our MSTO selection restricts the inferred stellar masses to a relatively narrow range of approximately $0.8$--$1.0~\msun$. Therefore, although the sample covers a broad range of chemical abundances, the variation in stellar mass is expected to have a limited impact on the binary-fraction measurements presented in this work.

\subsection{Upper Limit of the Period}

In our previous work, the upper limit on the orbital period was constrained by the mean free path for stellar encounters, which is appropriate for the dense environment of dwarf galaxies. In contrast, the Galactic field has a much lower stellar density, such that dynamical disruption through close encounters is negligible over a Hubble time. Therefore, in this work we do not impose an environment-dependent upper limit based on the mean free path. Instead, we adopt a fixed empirical upper bound of $\log P = 8$ (with $P$ in days), consistent with commonly used limits for wide binaries. This choice effectively sets the maximum orbital separation while avoiding unnecessary assumptions about environmental effects in the field.

It is important to distinguish this adopted upper bound from the period range directly constrained by the observations. To quantify the sensitivity of the DESI multi-epoch LOSV measurements, we perform mock tests in which binaries with periods longer than a given threshold are replaced by single stars while keeping the observing cadence and measurement uncertainties unchanged. We find that the recovered binary fraction begins to decrease when the threshold is shorter than $\sim10^4$ days, indicating that the present analysis is primarily sensitive to binaries with $P\lesssim10^4$ days. Under the adopted period distribution, binaries with $\log(P/{\rm day})<4$ account for approximately 37.2\% of the total binary population. Consequently, the observed LOSV variability directly constrains the short-period population, while the inferred total binary fraction is obtained by extrapolating to longer periods under the adopted orbital-parameter distributions.

\subsection{Color Dependence of Systematic Error Floor}
\label{ssec:syserr}
In our previous work, we modeled the systematic error floor in LOSV measurements as a single constant value, determined through a grid-based likelihood analysis. Subsequent investigations, however, indicate that this systematic error floor depends on stellar color. This dependence is driven by the instrumental characteristics of DESI: for bluer stars, the LOSV information is dominated by the $B$ and $R$ arms, where the wavelength calibration is less precise because of the relative scarcity of sky emission lines; for redder stars, the signal is dominated by the $Z$ arm, which benefits from more accurate calibration. As a result, bluer stars exhibit a systematically higher error floor than redder stars.

This color dependence also explains the initially counterintuitive program-level differences reported by \citet{2026OJAp....955260K}. They found an error floor of $\sim 1.6~{\rm km~s^{-1}}$ for the \textsc{dark} program, higher than the $\sim 1.0~{\rm km~s^{-1}}$ value for the \textsc{bright} program, despite the generally better observing conditions in \textsc{dark}. This difference is largely attributable to the target color distributions: \textsc{dark} targets are predominantly blue, whereas \textsc{bright} targets span a broader color range and extend to much redder colors.

Although the MSTO sample considered in this work spans a relatively narrow range in color, it lies predominantly in the blue regime, where the LOSV precision is most sensitive to small changes in color. Even modest variations in color can therefore lead to noticeable differences in the error floor. Motivated by this behavior, we adopt a color-dependent treatment of the systematic error floor in this study. Specifically, we divide the sample into bins in stellar color, and for each bin we independently determine the optimal systematic error floor and then assign the corresponding value to each star.

Following the approach in \citetalias{2026ApJ..1000...33Q}, we implement a grid-based strategy in which $\sigma_{\rm sys}$ is varied over a few discrete values. For each value, we recompute the LOSV variability distribution by adding Gaussian noise in quadrature to the observational uncertainties, and then perform the likelihood-based inference of the binary fraction. The preferred $\sigma_{\rm sys}$ in each color bin is identified by comparing the maximum log-likelihood values. 

\begin{figure}
\plotone{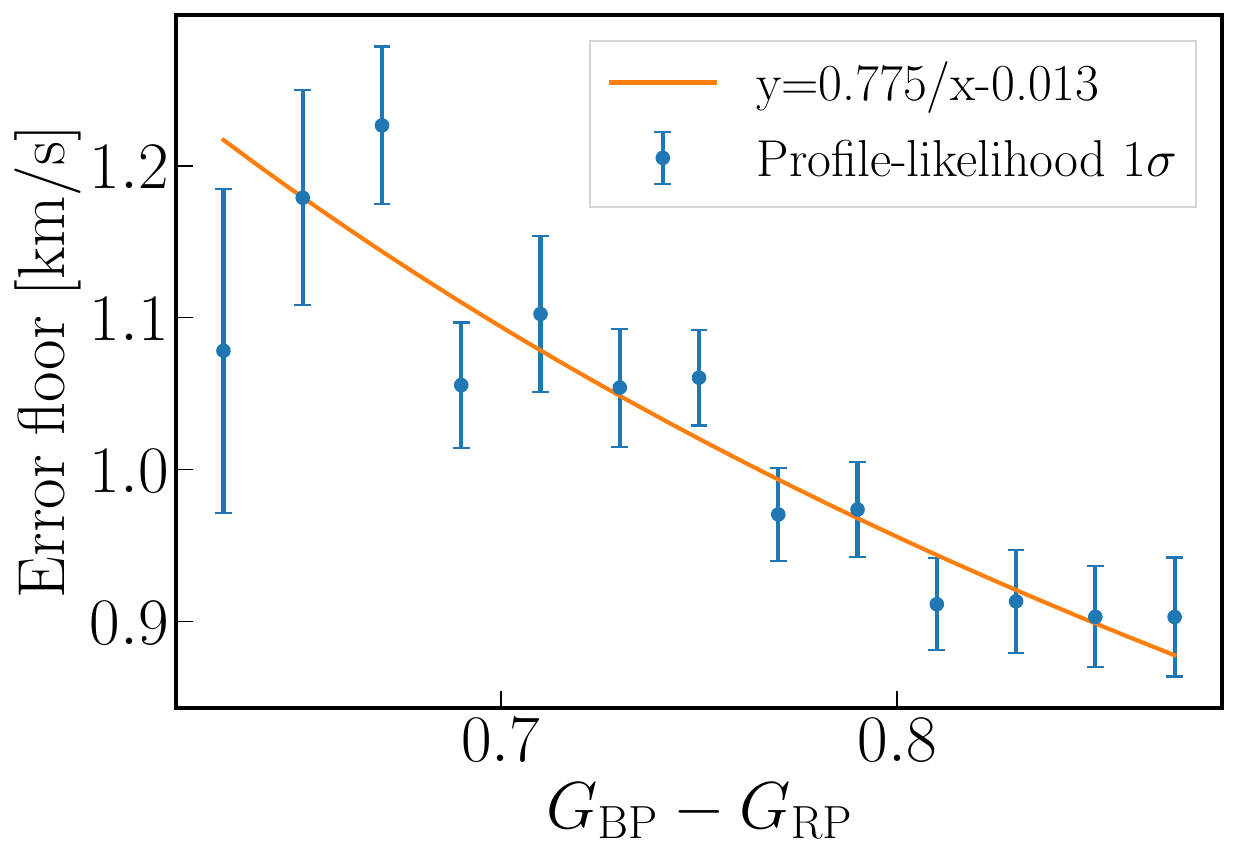}
\caption{LOSV error floor as a function of {\it Gaia} color, $G_{\rm BP}-G_{\rm RP}$. The sample is divided into color bins, and the error floor is determined independently in each bin from the profile likelihood. Points show the maximum-likelihood estimates, while the error bars indicate the $1\sigma$ confidence intervals defined by $\Delta\log\mathcal{L}=0.5$ relative to the likelihood maximum. The widths of these intervals depend on the statistical constraining power of each bin and are therefore generally larger for bins containing fewer stars. The orange curve shows the best-fitting relation, $\sigma_{\rm floor}=0.775/(G_{\rm BP}-G_{\rm RP})-0.013~{\rm km\,s^{-1}}$.
\label{fig:color_err}}
\end{figure}

To characterize the color dependence of the systematic error floor, we adopt the {\it Gaia} $G_{\rm BP}-G_{\rm RP}$ color, which provides a homogeneous color measurement for the full sample and is independent of the $g-r$ color used in the DESI target selection.
The resulting relation between $\sigma_{\rm sys}$ and $G_{\rm BP}-G_{\rm RP}$ is shown in Figure~\ref{fig:color_err}. In each color bin, we determine the maximum-likelihood value of $\sigma_{\rm sys}$ from the profile likelihood. The corresponding uncertainty is defined by the interval over which the log-likelihood decreases by $\Delta\log\mathcal{L}=0.5$ from its maximum, equivalent to the approximate $1\sigma$ confidence interval for a single fitted parameter. The widths of these intervals reflect the statistical constraining power of each color bin and are generally larger in bins containing fewer stars.
We find a clear trend in which the systematic error floor decreases toward redder colors. To facilitate its implementation in our simulations, we further fit this relation with a simple analytical function $y=a/x+b$, where $x=G_{\rm BP}-G_{\rm RP}$ is the {\it Gaia} color and $y=\sigma_{\rm sys}$ is the LOSV systematic error floor in km s$^{-1}$. Accordingly, $a=0.775\pm0.102$ km s$^{-1}$ mag and  $b=-0.013\pm0.132$  km s$^{-1}$. We assign the corresponding $\sigma_{\rm sys}$ to each star based on its color. We also tested linear and more flexible functional forms for this relation. However, because our MSTO sample spans only a limited color range, the resulting differences in the assigned $\sigma_{\rm sys}$ values are minor. 

\section{Results}
\label{sec:results}

\subsection{Binary Fraction Across Chemical-abundance Space}

\begin{figure*}
\plotone{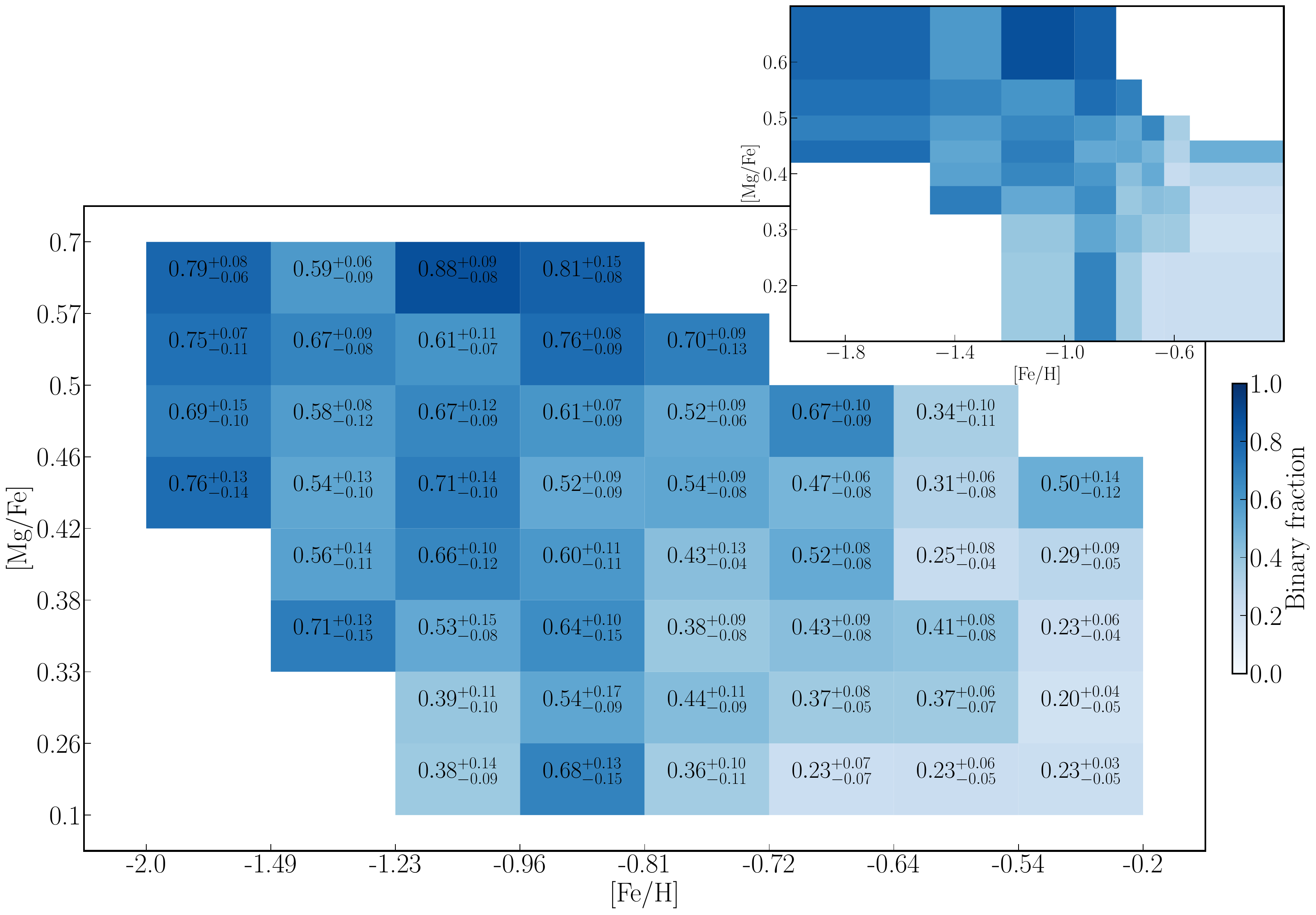}
\caption{
Binary fraction constraints across the [Fe/H]--[Mg/Fe] plane.
The stellar sample is divided into two-dimensional bins
in chemical-abundance space, with the grid boundaries defined in Section~\ref{ssec:chemgrid}. 
For each grid, the inset panel displays the PPD of the binary fraction, $f_b$, with the quoted value representing the median and corresponding $1\sigma$ credible interval. The background color indicates the median inferred binary fraction in each bin.
For visualization purposes, all grids in the main panel are displayed
with equal size, regardless of their actual extent in chemical-abundance space.
The upper-right panel therefore provides a
complementary view by projecting the inferred binary fractions onto their true locations in the [Fe/H]--[Mg/Fe] plane shown in Figure~\ref{fig:FeH_MgFe_grid}.
Overall, the binary fraction exhibits a clear dependence on [Fe/H], increasing systematically toward lower [Fe/H]. In contrast, the dependence on [Mg/Fe] is weaker and less uniform. Although some [Fe/H] intervals show a moderate increase in binary fraction with increasing [Mg/Fe], no universal monotonic trend is observed across the full abundance range.
\label{fig:bf_grid}}
\end{figure*}

The inferred binary fraction across the [Fe/H]–[Mg/Fe] plane is shown in Figure~\ref{fig:bf_grid}. Each cell displays the posterior median of $f_b$ and its 68\% credible interval. The quoted value above each distribution gives the median of the posterior together with its $1\sigma$ credible interval, while the background color represents the median inferred binary fraction. Since each grid has a different width in [Fe/H] or [Mg/Fe], the upper-right panel further presents the inferred binary fractions at their true locations, providing an overview of the distribution across [Fe/H]--[Mg/Fe] space.

A clear dependence of the binary fraction on [Fe/H] is observed. Metal-poor populations generally exhibit higher binary fractions, frequently reaching $f_b \gtrsim 0.5$ at $[\mathrm{Fe}/\mathrm{H}] \lesssim -0.8$, while more metal-rich populations tend to show substantially lower values of $f_b \sim 0.2$–0.4. This trend remains broadly consistent across multiple abundance bins despite the uncertainties reflected in the posterior distributions.

By comparison, the dependence on [Mg/Fe] is weaker. Although some [Fe/H] intervals exhibit moderately higher binary fractions towards larger [Mg/Fe], the distribution does not show a universal monotonic trend with [Mg/Fe]. The primary variation of the binary fraction therefore appears to be associated with [Fe/H].

\subsection{Dependence on [Mg/Fe]}
\label{ssec:mgfe}

Figure~\ref{fig:bf_grid} shows the joint variation of $f_b$ across the observed [Fe/H]–[Mg/Fe] plane. However, because [Fe/H] and [Mg/Fe] are correlated in the Galactic stellar population, a direct comparison across [Mg/Fe] bins can still retain differences in the underlying [Fe/H] distribution, even within the same broad [Fe/H] column. This makes it difficult to isolate an independent dependence of $f_b$ on [Mg/Fe], particularly because the number of high-[Mg/Fe] stars also decreases. We therefore construct [Mg/Fe]-selected subsamples whose [Fe/H] distributions are matched to a common target distribution.

Specifically, the sample is divided into six non-overlapping [Mg/Fe] slices over the range of $0.1 \leq [\mathrm{Mg}/\mathrm{Fe}] \leq 0.7$, each with bin width of $[\mathrm{Mg}/\mathrm{Fe}]=0.1$. Within each slice, we resample the stars to match a Gaussian target distribution in [Fe/H]. The target distribution has a standard deviation of $\sigma_{[\mathrm{Fe}/\mathrm{H}]}=0.1$ dex, with centers ranging from $[\mathrm{Fe}/\mathrm{H}]=-1.8$ to $-0.4$ dex in steps of 0.1 dex. The adopted width of 0.1 dex represents a compromise between minimizing residual [Fe/H] variation and maintaining sufficient sample size. Owing to the limited number of stars in some regions of chemical-abundance space, not every combination of [Fe/H] and [Mg/Fe] can have enough stars. Sampling is performed without replacement within each [Fe/H] and [Mg/Fe] combination, though the same star may appear in different realizations or in different combinations. Compared with applying narrow [Fe/H] cuts, the Gaussian resampling substantially increases the number of usable stars in each abundance bin while maintaining nearly identical [Fe/H] distributions among the different [Mg/Fe] slices.

This Gaussian-matching procedure allows different [Mg/Fe] populations to be compared at approximately the same [Fe/H] distribution while retaining more stars than would be possible with narrow, non-overlapping [Fe/H] cuts. Because the resampled subsamples span abundance ranges that differ from the two-dimensional bins used in Figure~\ref{fig:bf_grid}, the measurements are independently inferred again. Regions that do not meet the minimum sample-size requirement in Figure~\ref{fig:bf_grid} can therefore still contribute stars to a sufficiently matched subsample. The broader and partially overlapping target distributions, together with repeated resampling, also naturally produce a smoother representation than the direct two-dimensional abundance map.

\begin{figure*}
\plotone{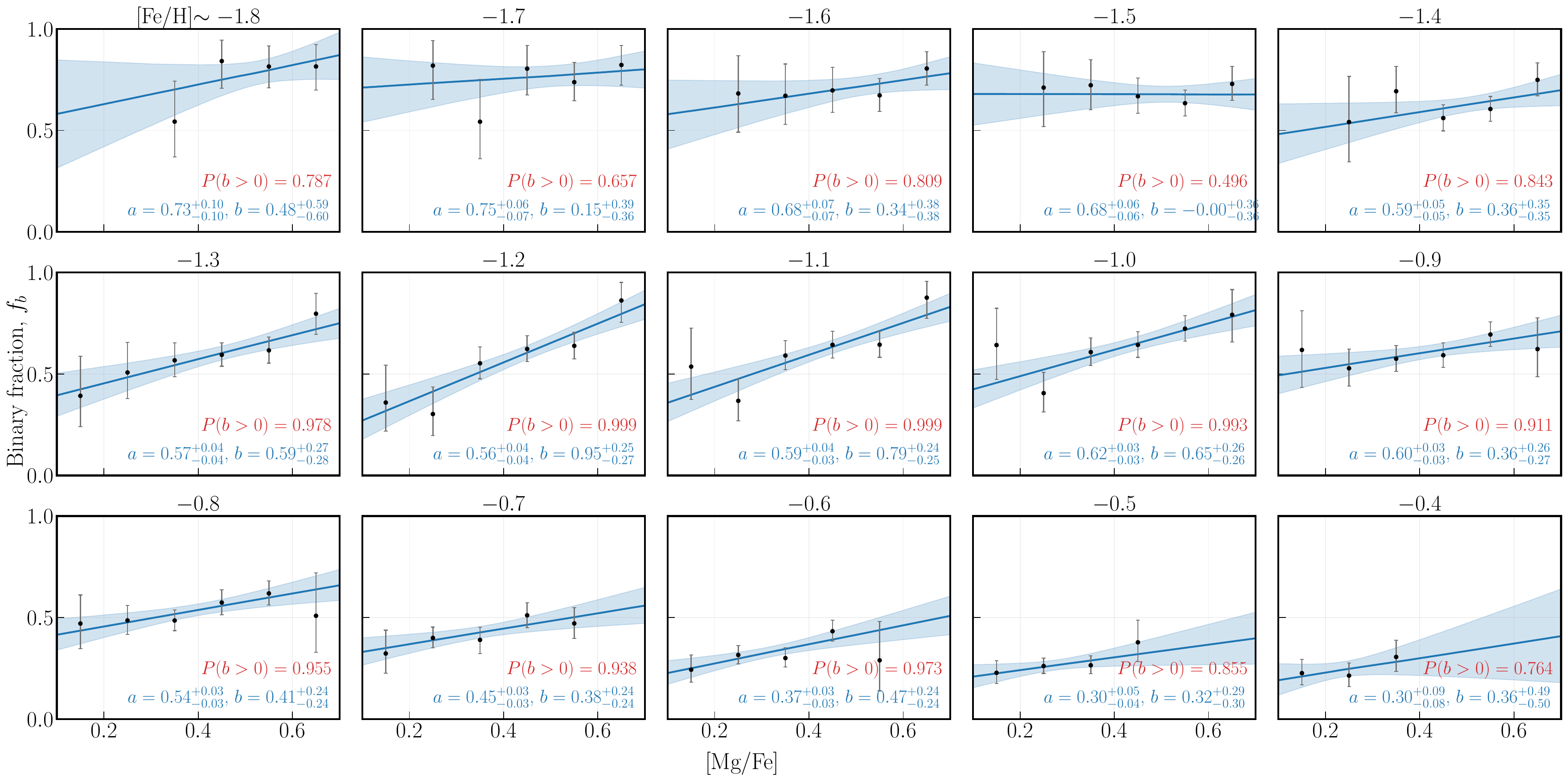}
\caption{
Binary fraction as a function of [Mg/Fe] at fixed [Fe/H]. The sample is divided into six non-overlapping [Mg/Fe] slices. Within each slice, Monte Carlo resampling is used to construct [Fe/H]-matched subsamples following Gaussian distributions in [Fe/H] with $\sigma_{[\mathrm{Fe}/\mathrm{H}]}=0.1$ dex and centers ranging from $-1.8$ to $-0.4$ dex. Black points show the inferred binary fractions, with error bars indicating the corresponding $1\sigma$ uncertainties. The blue lines show the posterior median linear relations, $f_b=a+b([\mathrm{Mg}/\mathrm{Fe}]-0.4)$, and the blue shaded regions denote their $1\sigma$ credible intervals. The inferred intercept $a$, slope $b$, and posterior probability $P(b>0)$ are presented in each panel.
\label{fig:bf_feh_mgfe_slices}}
\end{figure*}

\begin{figure}
\plotone{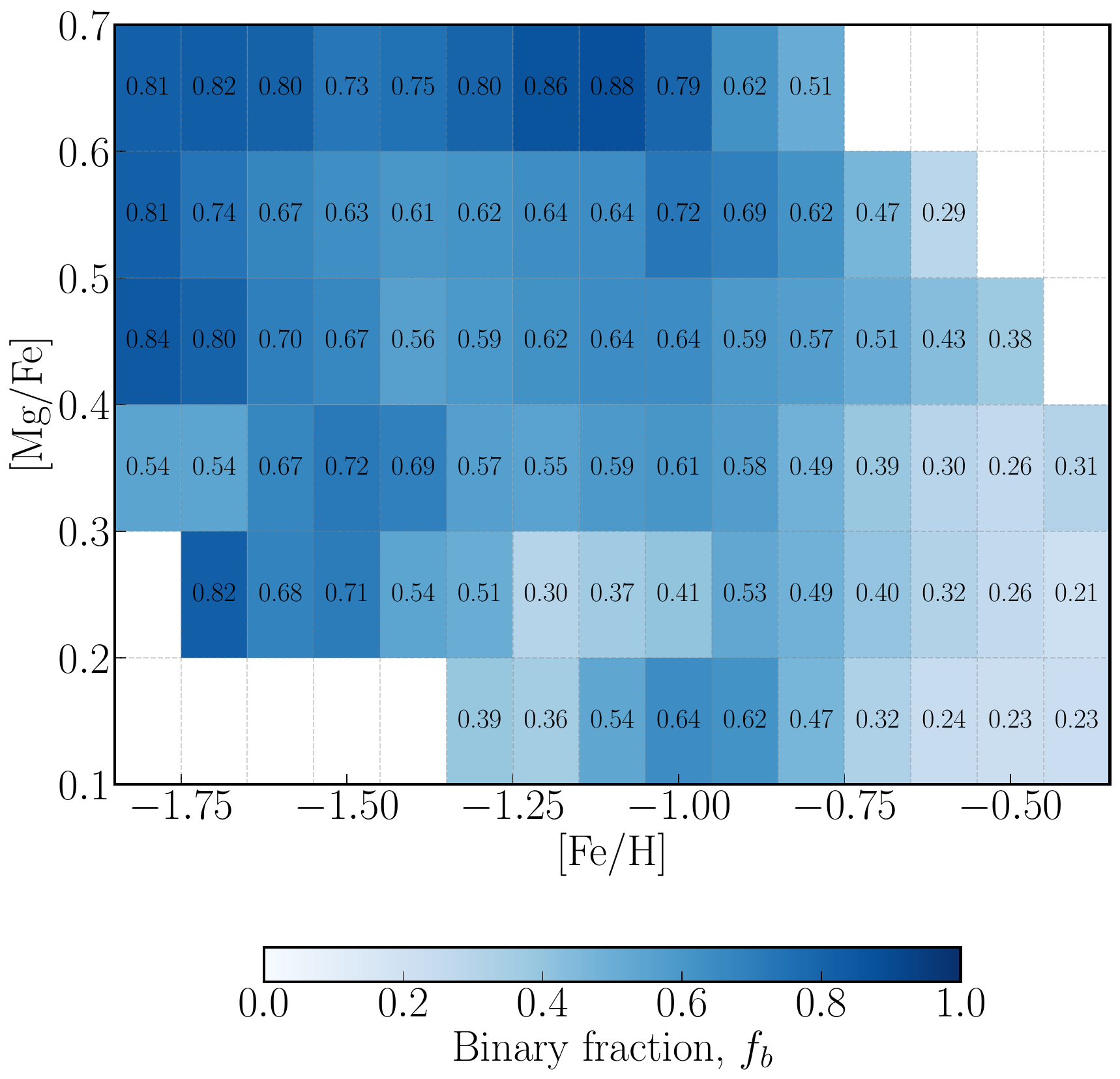}
\caption{
Summary of the binary fraction across the [Fe/H]–[Mg/Fe] plane after [Fe/H] matching. Each cell shows the binary fraction obtained from the Monte Carlo [Fe/H]-matched subsamples shown in Figure~\ref{fig:bf_feh_mgfe_slices}.
The map highlights the dominant increase of binary fraction toward lower [Fe/H]. Some fluctuations are present in the sparse regions of the abundance plane, particularly toward the metal-poor and low-[Mg/Fe] corner, where the limited number of available stars makes the resampling more sensitive to individual measurements and increases the statistical uncertainty.
\label{fig:bf_feh_mgfe_map}}
\end{figure}

The resulting binary fractions are shown in Figure~\ref{fig:bf_feh_mgfe_slices}. By construction, the measurements in each panel have nearly identical [Fe/H] distributions but span different [Mg/Fe] ($x$-axis), substantially reducing the correlation between [Fe/H] and [Mg/Fe]. For each [Fe/H] interval, we fit a linear relation, 
\begin{align}
f_b = a + b([\mathrm{Mg}/\mathrm{Fe}] - 0.4)    
\end{align}
using the posterior binary-fraction distributions of all abundance bins. The blue line shows the posterior median relation, while the shaded region represents the corresponding $1\sigma$ credible interval. The inferred intercept $a$, slope $b$, and the posterior probability that the slope is positive, $P(b>0)$, are listed in each panel.

A clear dependence on [Fe/H] remains evident. The fitted intercept, $a$, which represents the binary fraction at the pivot point $[\mathrm{Mg}/\mathrm{Fe}]=0.4$, increases systematically from the metal-rich to the metal-poor panels, consistent with the trend identified in the two-dimensional abundance maps (Figure~\ref{fig:bf_grid}). At fixed [Fe/H], the dependence on [Mg/Fe] is much weaker. Nevertheless, for relatively metal-rich populations ($[\mathrm{Fe}/\mathrm{H}] \gtrsim -1.3$), the binary fraction shows a modest increase with increasing [Mg/Fe]. The only exception is the most metal-rich bin ($[\mathrm{Fe}/\mathrm{H}]\sim-0.4$), where the constraint is weaker because of the limited number of stars. The significance of the positive slope is quantified by the posterior probability $P(b>0)$, where $P(b>0)=0.84$ and $0.975$ correspond approximately to 1 and 2$\sigma$ evidence, respectively. Most panels with $[\mathrm{Fe}/\mathrm{H}] \gtrsim -1.3$ have $P(b>0)\gtrsim0.975$, indicating moderate to strong evidence for a positive correlation. In contrast, for more metal-poor populations ($[\mathrm{Fe}/\mathrm{H}] \lesssim -1.3$), the inferred slopes are consistent with zero within the uncertainties, with substantially lower values of $P(b>0)$. The broader credible intervals in these panels primarily reflect the smaller numbers of stars available at the metal-poor end.

To facilitate comparisons across the entire chemical abundance plane, Figure~\ref{fig:bf_feh_mgfe_map} summarizes the binary fraction for each [Fe/H]-matched abundance bin. The map confirms that [Fe/H] is the primary parameter governing the binary fraction in the Galactic field, while any dependence on [Mg/Fe] is secondary and varies with [Fe/H]. Although a positive correlation with [Mg/Fe] is visible in several relatively metal-rich bins, no universal monotonic trend is observed over the full [Fe/H] range. Some apparent fluctuations are seen in the sparsely populated regions of the abundance plane, especially toward the low-[Fe/H] and low-[Mg/Fe] corner. These variations are likely amplified by the limited number of stars available for the [Fe/H]-matched resampling, where repeated sampling of a small number of stars can increase the sensitivity to individual velocity-variable systems. Therefore, these fluctuations should be interpreted with caution and do not alter the dominant chemical abundance dependence observed across the full range.

\subsection{Dependence on Galactic Position}
\label{ssec:position}

To investigate whether the binary fraction exhibits additional dependence on the Galactic environment after accounting for its correlation with stellar chemical abundances, we examine its variation with Galactic position. Using the Galactocentric coordinates of the MSTO sample, we explore trends with both Galactocentric radius, $r_{\rm GC}$, and vertical distance from the Galactic plane, $|z|$. The radial dependence probes possible variations associated with large-scale Galactic structure, while the vertical dependence provides sensitivity to different Galactic populations, including the thin disk, thick disk, and halo. A detailed comparison among these populations using combined chemical and kinematic information is presented in the next subsection. Since chemical abundance and Galactic position are strongly correlated in the Milky Way, this analysis complements the abundance-space results discussed above and helps determine whether the observed binary fraction variations are primarily associated with chemical abundances or present-day Galactic environment.

\begin{figure}
\plotone{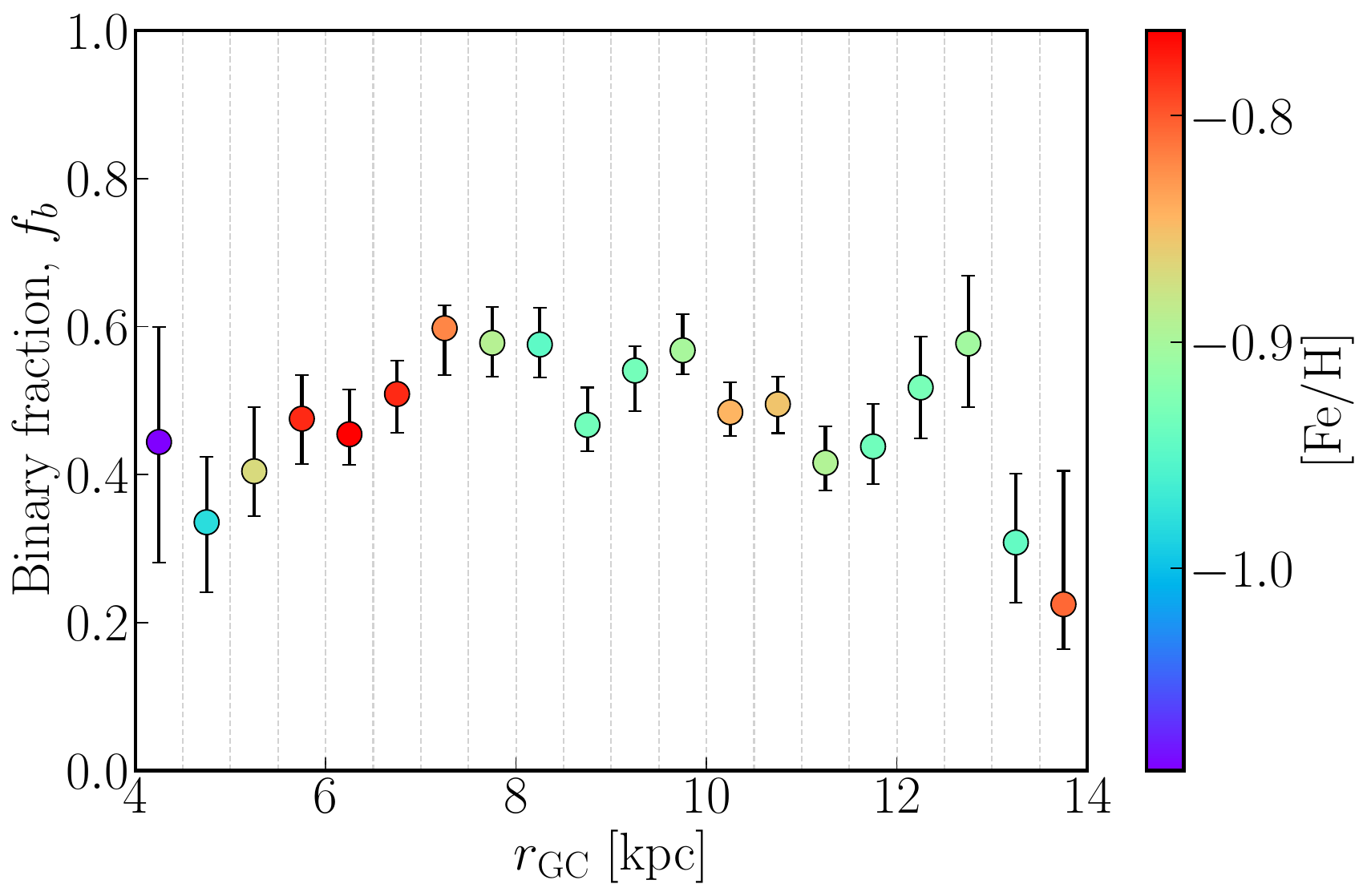}
\caption{
Binary fraction as a function of Galactocentric radius, $r_{\rm GC}$. The stellar sample is divided into 0.5 kpc radial bins. The binary fraction is inferred independently in each bin, with error bars indicating the $1\sigma$ credible interval. The color scale represents the mean [Fe/H] of each bin. No significant radial dependence of the binary fraction is observed over the range covered by the sample.
\label{fig:bf_R_slices}}
\end{figure}

Figure~\ref{fig:bf_R_slices} shows the inferred binary fraction as a function of Galactocentric radius, $r_{\rm GC}$, with each point colored by the mean [Fe/H] of the corresponding bin. No significant trend is observed over the radial range covered by our sample. The inferred binary fractions remain broadly consistent within their uncertainties, with only moderate bin-to-bin fluctuations.

In addition to the instantaneous Galactocentric radius, we investigate the dependence of the binary fraction on the vertical component of the angular momentum, $L_z$. Since $L_z$ is approximately conserved in an axisymmetric Galactic potential, it provides a more stable tracer of stellar orbital properties than the present-day spatial position.

\begin{figure}
\plotone{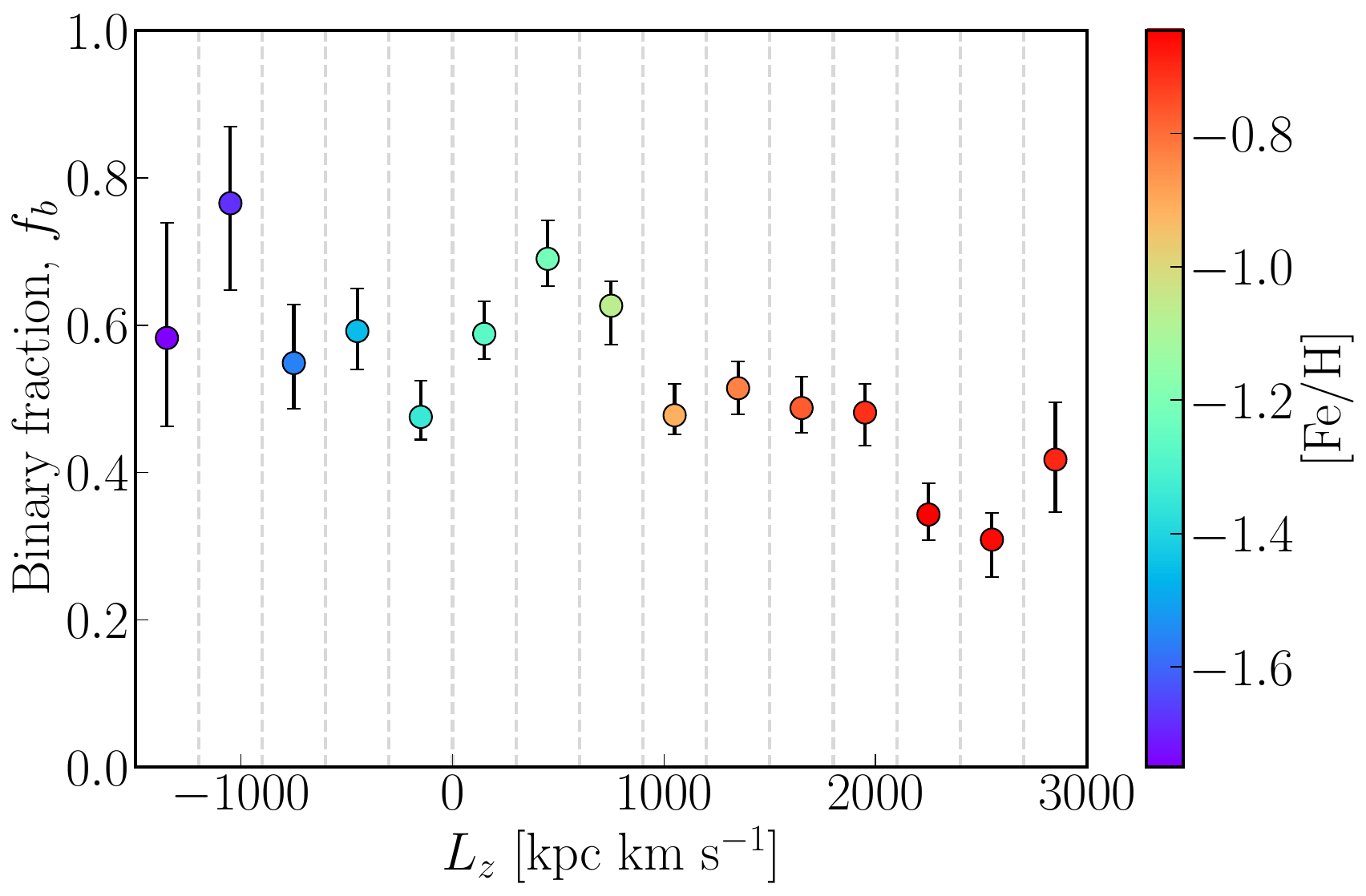}
\plotone{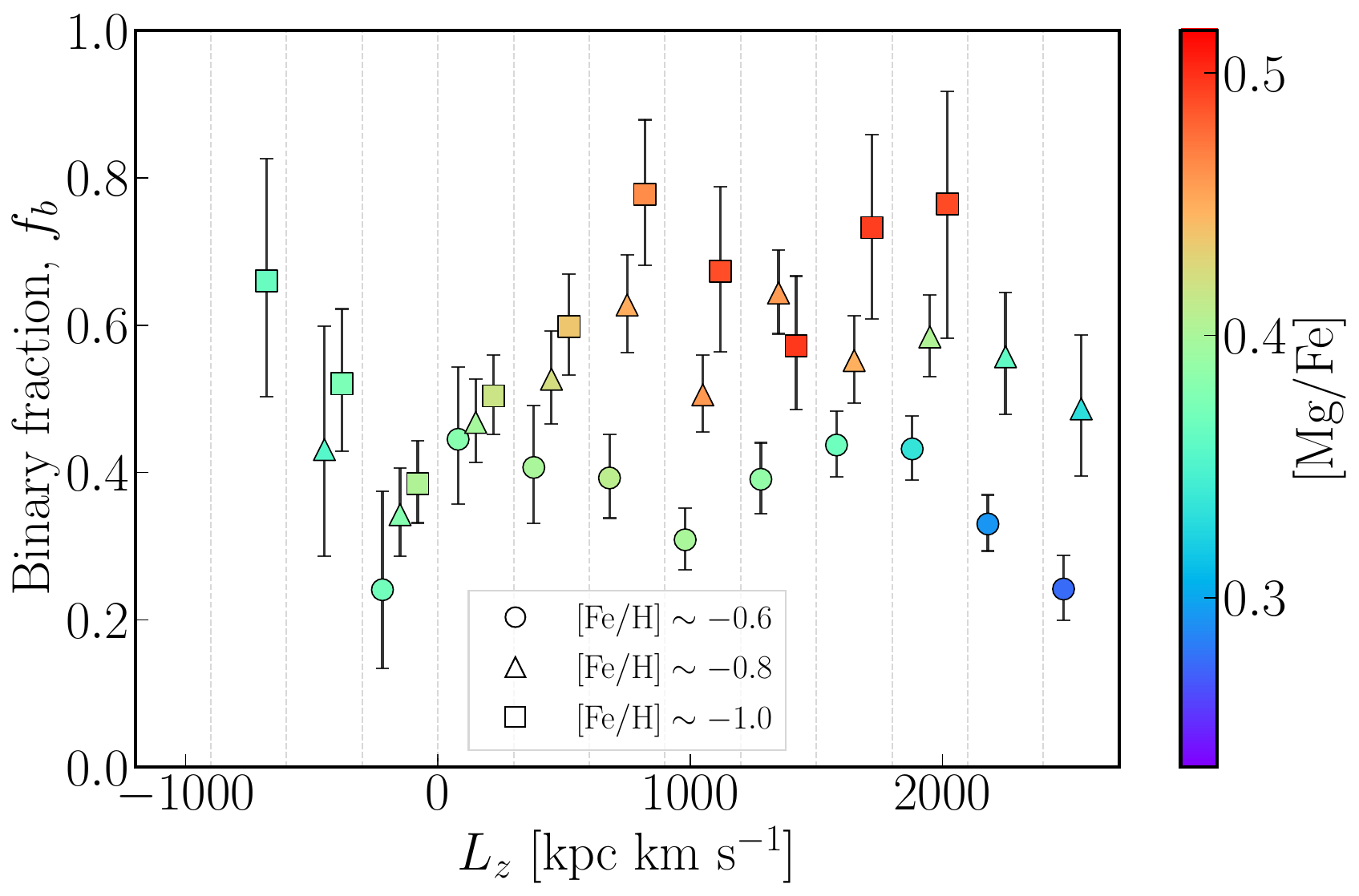}
\caption{
\textbf{Upper panel:} Binary fraction as a function of the vertical component of angular momentum, $L_z$. The stellar sample is divided into bins of $L_z$, and the binary fraction is inferred independently in each bin. Error bars indicate the corresponding $1\sigma$ credible intervals. The color scale represents the mean [Fe/H] of each bin. The apparent decline of the binary fraction toward larger positive $L_z$ follows the variation of the [Fe/H] distribution.\\
\textbf{Lower panel:} Binary fraction as a function of $L_z$, after controlling for [Fe/H]. Different symbols denote subsamples drawn from Gaussian [Fe/H] distributions centered at $-0.6$, $-0.8$, and $-1.0$ dex, each with $\sigma_{[\mathrm{Fe}/\mathrm{H}]}=0.1$ dex. Symbols corresponding to different [Fe/H] subsamples are slightly offset horizontally for clarity. The color scale represents the mean [Mg/Fe] of each subsample. After controlling for [Fe/H], no significant dependence on $L_z$ is observed, while the remaining variations are consistent with the [Mg/Fe] abundance dependence identified above.
\label{fig:bf_Lz_slices}}
\end{figure}

The upper panel of Figure~\ref{fig:bf_Lz_slices} shows the inferred binary fraction as a function of $L_z$, with the points colored by the mean [Fe/H] of each bin. The binary fraction shows a broad decline toward larger positive $L_z$. However, the [Fe/H] distribution varies systematically with angular momentum, with lower-$L_z$ stars being more metal-poor on average. This suggests that the apparent $f_b$--$L_z$ relation may largely reflect the underlying [Fe/H] distribution rather than an independent dependence on orbital angular momentum.

To separate the effects of orbital properties and chemical abundance, we repeat the analysis after controlling for [Fe/H] using the Monte Carlo resampling procedure described in Section~\ref{ssec:mgfe}. Within each $L_z$ bin, we construct subsamples with matched Gaussian distributions centered at $[\mathrm{Fe}/\mathrm{H}]=-0.6$, $-0.8$, and $-1.0$, each with $\sigma_{[\mathrm{Fe}/\mathrm{H}]}=0.1$ dex.

The resulting measurements are shown in the lower panel of Figure~\ref{fig:bf_Lz_slices}. After matching the [Fe/H] distributions, the broad variation of $f_b$ with $L_z$ is substantially weakened. At a given $L_z$, the more metal-poor subsamples generally retain higher binary fractions, consistent with the dominant [Fe/H] dependence identified above. The [Fe/H]-controlled measurements do not show a simple monotonic dependence on $L_z$; however, a mild suppression of the binary fraction is visible around $L_z\sim0$, particularly for the metal-poor subsamples. This residual feature may indicate differences among dynamically distinct populations, although the current uncertainties prevent a strong constraint. The remaining variations also broadly follow changes in the mean [Mg/Fe], suggesting that part of the residual structure may be associated with the secondary chemical dependence identified in Section~\ref{ssec:mgfe}.

\begin{figure}
\plotone{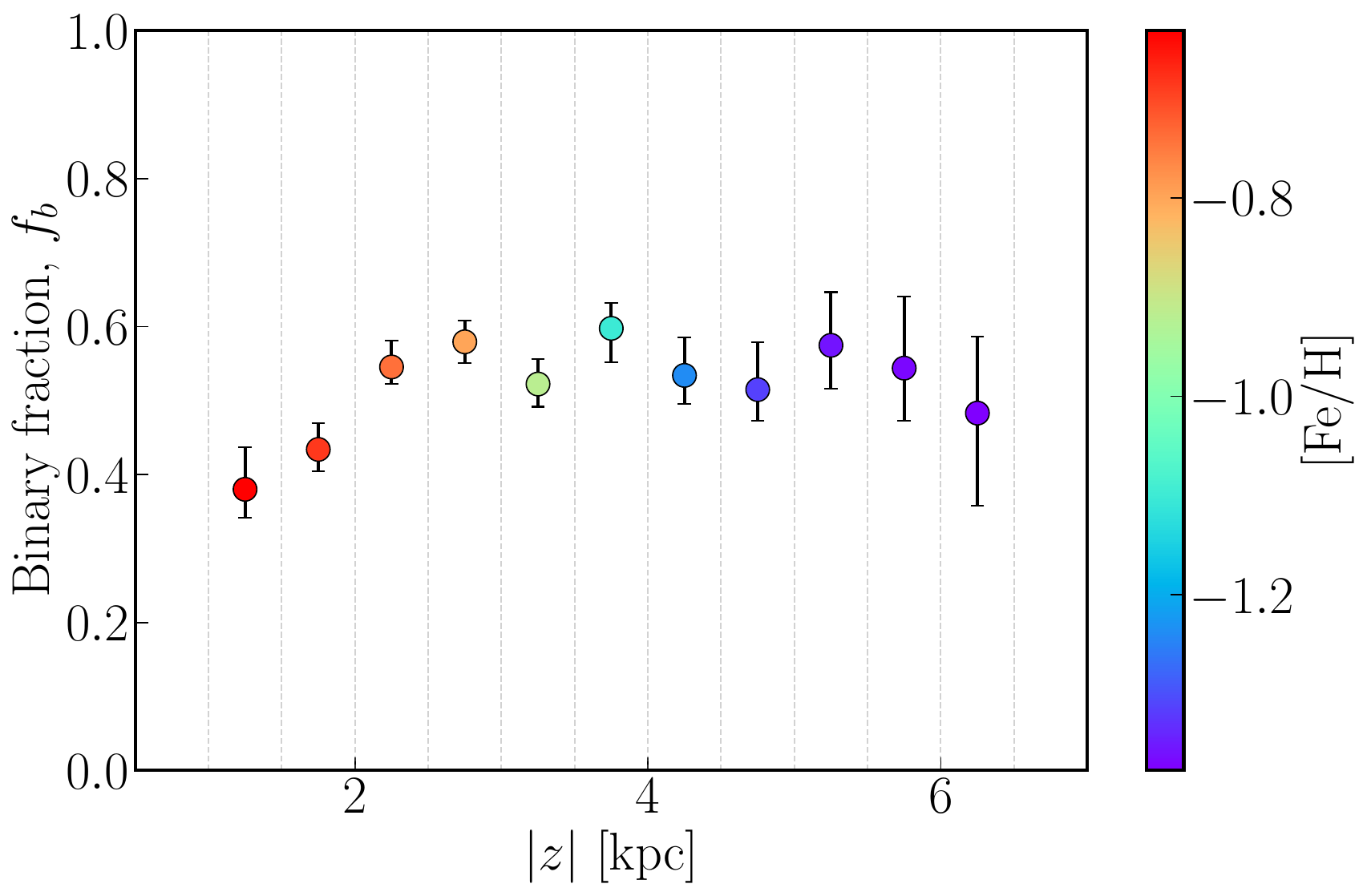}
\plotone{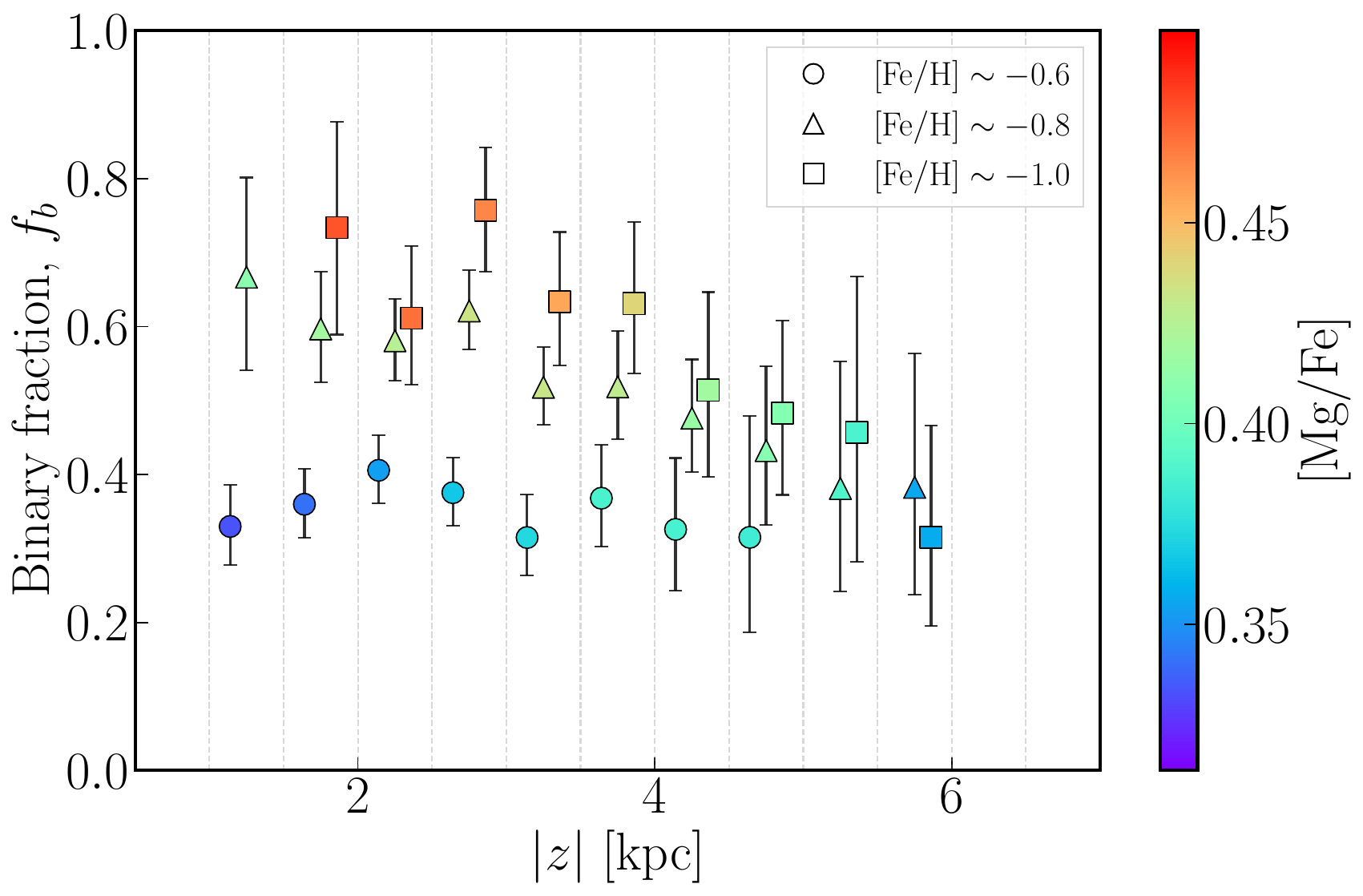}
\caption{
\textbf{Upper panel:} Binary fraction as a function of Galactic vertical distance, $|z|$, before controlling for [Fe/H]. Points show the inferred binary fraction in each $|z|$ bin, with error bars indicating the corresponding $1\sigma$ credible intervals. The color scale represents the mean [Fe/H] of each bin. The apparent dependence on $|z|$ is accompanied by a systematic variation in [Fe/H].\\
\textbf{Lower panel:} Similar to Figure~\ref{fig:bf_Lz_slices}. Binary fraction as a function of $|z|$, after controlling for [Fe/H]. Different symbols show subsamples drawn from Gaussian [Fe/H] distributions centered at $-0.6$, $-0.8$, and $-1.0$ dex with $\sigma=0.1$ dex. The color scale indicates the mean [Mg/Fe] of each subsample. After controlling for [Fe/H], only weak residual variations with $|z|$ remain, while the dominant trend is associated with [Mg/Fe].
\label{fig:bf_Z_slices}}
\end{figure}

We next investigate the dependence on vertical distance from the Galactic plane. The upper panel of Figure~\ref{fig:bf_Z_slices} shows the inferred binary fraction as a function of $|z|$, with the colors indicating the mean [Fe/H] of each bin. A broad increase of the binary fraction toward larger $|z|$ is visible, although the trend is not strictly monotonic. Since stars at larger vertical distances are generally more metal-poor, the apparent $f_b$--$|z|$ relation may similarly reflect the underlying [Fe/H] gradient rather than a direct dependence on Galactic height.

To further examine whether the apparent $f_b$--$|z|$ relation is driven by Galactic height or by the underlying [Fe/H] distribution, we repeat the [Fe/H]-controlled analysis using the same resampling procedure. Three independent realizations of the resampled subsamples are generated to evaluate the robustness of the inferred trends. The resulting measurements are presented in the lower panel of Figure~\ref{fig:bf_Z_slices}.

After controlling for [Fe/H], the broad increase of $f_b$ toward larger $|z|$ almost disappears, though the two lower [Fe/H] subsamples show residual variations with the height. At fixed $|z|$, more metal-poor subsamples generally retain higher binary fractions than more metal-rich subsamples, consistent with the dominant [Fe/H] dependence identified above. Some residual variations remain, particularly for the $[\mathrm{Fe}/\mathrm{H}]\sim-0.8$ and $-1.0$ subsamples, where the binary fraction shows a weak decrease toward larger $|z|$. These residual variations broadly follow the mean [Mg/Fe] distribution, suggesting that part of the remaining scatter may be related to differences in chemical abundance beyond [Fe/H].

The results above indicate that the binary fraction is more strongly associated with stellar chemical abundances, especially [Fe/H], than with present-day Galactic position. Although weak residual variations remain, they are comparable to the level expected from differences in other chemical abundances such as [Mg/Fe]. Galactic position alone does not uniquely identify the underlying stellar populations, since the thin disk, thick disk, and stellar halo overlap substantially in both spatial and chemical parameter spaces. Therefore, in the following subsection, we classify the sample into Galactic components using combined chemical and kinematic information and compare the inferred binary fractions among populations with different formation histories.

\subsection{Binary Fractions Across Galactic Components}
\label{ssec:components}

In this subsection, we further classify the MSTO sample into five chemically and kinematically selected Galactic components: the thin disk, thick disk, Gaia--Sausage--Enceladus (GSE), Splash population and the remaining stellar halo. The classification combines chemical abundances with instantaneous phase-space quantities, including $v_r$, $v_\phi$, $v_{\rm pec}$, $L_z$, and orbital energy, $E$. 

Before dividing stars to different Galactic components, we apply additional quality cuts to ensure reliable distances and chemical abundances. We require the fractional distance uncertainty to satisfy
$\sigma_d/d < 0.5$, together with
$\sigma_{[\mathrm{Fe}/\mathrm{H}]} < 0.3$ dex
and
$\sigma_{[\mathrm{Mg}/\mathrm{Fe}]} < 0.35$ dex.
Because we rely on distance and chemical abundance measurements to classify stars into different Galactic components, we select this subset of sample with more precise measurements for robust analysis in this subsection. 

The classification of stars into different Galactic components is based on [Fe/H] and [Mg/Fe], Galactocentric radial and azimuthal velocities, $v_r$ and $v_\phi$, peculiar velocity\footnote{This is the velocity with respect to the Sun, so the average azimuthal velocity component ($v_{\phi_0}=232~{\rm km\,s^{-1}}$) has been subtracted.}, $v_{\rm pec}=\sqrt{v_r^2+v_z^2+(v_\phi-v_{\phi_0})^2}$, vertical angular momentum, $L_z$, and orbital energy, $E$. The angular momentum and energy are calculated in the Galactic potential adopted in this work. We use instantaneous phase-space quantities and do not require orbital integration. All selections are applied after the common astrometric, spectroscopic, and phase-space quality cuts above. 

We separate thick and thin disk stars according to the following criterion in chemical abundance space
\begin{equation}
[\mathrm{Mg}/\mathrm{Fe}]=0.18-0.25[\mathrm{Fe}/\mathrm{H}].
\label{eq:alpha}
\end{equation}
This boundary was defined empirically by the [Mg/Fe]--[Fe/H] distribution in Figure~\ref{fig:FeH_MgFe_grid}, shown by the black dashed line. Stars below and above this relation are low-[Mg/Fe] and high-[Mg/Fe] populations.

Thin-disk candidates are selected from the low-[Mg/Fe] sequence and are further required to satisfy
\begin{align}
[\mathrm{Fe}/\mathrm{H}]&>-0.8, \nonumber \\
v_\phi&>180~{\rm km\,s^{-1}},\\
v_{\rm pec}&<100~{\rm km\,s^{-1}}. \nonumber
\end{align}
These criteria select metal-rich, rotationally supported stars with relatively cold kinematics.

Thick-disk candidates are selected from the high-[Mg/Fe] sequence and further satisfy
\begin{align}
[\mathrm{Fe}/\mathrm{H}]&>-1.2,\nonumber \\
100<v_\phi&<220~{\rm km\,s^{-1}},\\
v_{\rm pec}&<180~{\rm km\,s^{-1}}. \nonumber
\end{align}
The broader velocity limits account for the lower mean rotation and larger velocity dispersion of the thick disk relative to the thin disk.

GSE candidates are identified through their low [Fe/H], weak net azimuthal rotation, low angular momenta, and strongly radial or relatively high-energy kinematics. We require
\begin{align}
[\mathrm{Fe}/\mathrm{H}]&<-0.7,\nonumber \\
|v_\phi|&<80~{\rm km\,s^{-1}},\\
|L_z|&<500~{\rm kpc\,km\,s^{-1}},\nonumber
\end{align}
together with either
\begin{equation}
|v_r|>100~{\rm km\,s^{-1}}\quad \mathrm{or} \quad
E>-1.6\times10^{5}~{\rm km^2\,s^{-2}}.
\end{equation}
The radial-velocity criterion selects stars on strongly radial trajectories, while the energy criterion retains low-angular-momentum GSE candidates whose instantaneous radial velocities may be small at their current orbital phases.

Splash candidates are identified as chemically disk-like stars with halo-like kinematics. Without explicitly separating this population, many Splash stars would be classified as halo members based solely on their kinematics, contaminating the other halo sample. We therefore identify Splash candidates separately and remove them from the halo category. We require
\begin{align}
[\mathrm{Fe}/\mathrm{H}]&>-0.7,\nonumber \\
|v_\phi|&<100~{\rm km\,s^{-1}},\\
|L_z|&<900~{\rm kpc\,km\,s^{-1}},\nonumber\\
v_{\rm pec}&>180~{\rm km\,s^{-1}}.\nonumber
\end{align}
These criteria isolate relatively metal-rich stars with weak rotation and dynamically hot kinematics. Such properties are characteristic of the Splash population, which is commonly interpreted as an early in-situ disk population that was dynamically heated during the formation of the inner stellar halo. The combined chemical and kinematic selection also reduces contamination from metal-rich thin-disk stars and accreted populations. 

The remaining halo category is constructed from stars that are not assigned to GSE or either disk component and satisfy at least one of the following conditions:
\begin{equation}
[\mathrm{Fe}/\mathrm{H}]<-1.0,
\end{equation}
\begin{equation}
v_{\rm pec}>180~{\rm km\,s^{-1}},
\end{equation}
or
\begin{equation}
|L_z|<700~{\rm kpc\,km\,s^{-1}}
\quad\mathrm{and}\quad
E>-1.8\times10^{5}~{\rm km^2\,s^{-2}}.
\end{equation}
This category includes halo stars that do not satisfy the more restrictive GSE selection, including stars with prograde, retrograde, and weakly rotating kinematics.

The adopted chemical and kinematic criteria for the five Galactic components are summarized in Table~\ref{tab:component_selection}. Because several of these criteria overlap, we impose a hierarchical assignment to obtain mutually exclusive component samples. GSE candidates are identified first, followed by the splash, thin- and thick-disk samples, and the remaining halo-like stars are assigned to the other-halo category. Stars that satisfy none of these selections are left unclassified and are excluded from the component-by-component comparison. The resulting groups should therefore be interpreted as relatively high-purity analysis samples rather than complete representations of the corresponding Galactic populations.

\begin{table*}
\centering
\caption{Chemical and kinematic criteria used to define Galactic components.}
\label{tab:component_selection}
\begin{tabular}{lcccccc}
\hline
Component &
Chemical selection &
$v_\phi$ &
$v_r$ &
$v_{\rm pec}$ &
$L_z$ &
$E$ \\
 &
 &
$(\mathrm{km\,s^{-1}})$ &
$(\mathrm{km\,s^{-1}})$ &
$(\mathrm{km\,s^{-1}})$ &
$(\mathrm{kpc\,km\,s^{-1}})$ &
$(\mathrm{km^2\,s^{-2}})$ \\
\hline
Thin disk &
$[\mathrm{Fe}/\mathrm{H}]>-0.8$, low-[Mg/Fe]\tablenotemark{a} &
$>180$ &
$\cdots$ &
$<100$ &
$\cdots$ &
$\cdots$ \\

Thick disk &
$[\mathrm{Fe}/\mathrm{H}]>-1.2$, high-[Mg/Fe]\tablenotemark{a} &
$100$--$220$ &
$\cdots$ &
$<180$ &
$\cdots$ &
$\cdots$ \\

GSE &
$[\mathrm{Fe}/\mathrm{H}]<-0.7$ &
$|v_\phi|<80$ &
$|v_r|>100$\tablenotemark{b} &
$\cdots$ &
$|L_z|<500$ &
$>-1.6\times10^5$\tablenotemark{b} \\

Splash &
$[\mathrm{Fe}/\mathrm{H}]>-0.7$ &
$|v_\phi|<100$ &
$\cdots$ &
$>180$ &
$|L_z|<900$ &
$\cdots$ \\

Other halo &
$[\mathrm{Fe}/\mathrm{H}]<-1.0$\tablenotemark{c} &
$\cdots$ &
$\cdots$ &
$>180$\tablenotemark{c} &
$|L_z|<700$\tablenotemark{c} &
$>-1.8\times10^5$\tablenotemark{c} \\
\hline
\end{tabular}

\tablenotetext{a}{The low- and high-[Mg/Fe] populations are separated using the empirical boundary
$[\mathrm{Mg}/\mathrm{Fe}]=-0.25[\mathrm{Fe}/\mathrm{H}]+0.18$.}
\tablenotetext{b}{A GSE candidate must satisfy the low-$L_z$ criterion and either the radial-velocity or energy criterion.}
\tablenotetext{c}{The other-halo selection requires at least one of the listed chemical-abundance, velocity, or joint $L_z$--$E$ conditions and is applied only after removing stars assigned to the other components.}
\end{table*}

The numbers of stars assigned to the different components are summarized in Table~\ref{tab:component_counts}. A total of 55,817 stars satisfy the additional distance and abundance-quality requirements. Of these, 6,776 stars are classified as thin-disk members; the relatively small size of this subsample reflects the distance range and Galactic-latitude coverage of our sample. The thick-disk and GSE samples contain 14,377 and 13,947 stars, respectively, while the Splash and other-halo components comprise 1,971 and 13,375 stars. The remaining 5,371 stars do not satisfy any of the adopted component definitions and are therefore retained as unclassified. These relative fractions reflect the spatial, chemical, and kinematic selection of the present our sample and should not be interpreted as the intrinsic population fractions of the Milky Way.

\begin{table}
\centering
\caption{Numbers of Stars Assigned to the Galactic Components}
\label{tab:component_counts}
\begin{tabular}{lrr}
\hline
Component & $N$ & Fraction (\%) \\
\hline
Thin disk   &  6,776 & 12.1 \\
Thick disk  & 14,377 & 25.8 \\
GSE         & 13,947 & 25.0 \\
Splash      &  1,971 & 3.5 \\
Other halo  & 13,375 & 24.0 \\
Unclassified & 5,371 & 9.6 \\
All classified components & 50,446 & 90.4 \\
Quality-controlled sample  & 55,817 & 100.0 \\
\hline
Failed quality cuts & 1,830 & \nodata \\
Parent sample       & 57,647 & \nodata \\
\hline
\end{tabular}
\end{table}

We then apply the same binary-fraction inference procedure to each component. For the Splash population, the binary fraction inferred from the full sample is $f_b=0.37^{+0.06}_{-0.05}$. However, the number of Splash stars is too small to obtain reliable constraints after further dividing the sample by chemical abundance. Therefore, the abundance-dependent analysis is restricted to the other four components. Since these four components occupy different regions of chemical-abundance space, a direct comparison of their overall binary fractions may remain affected by their different [Fe/H] distributions. We therefore examine both the component-averaged measurements and, where the sample size permits, comparisons performed within restricted or matched [Fe/H] intervals.

\begin{figure*}
\plotone{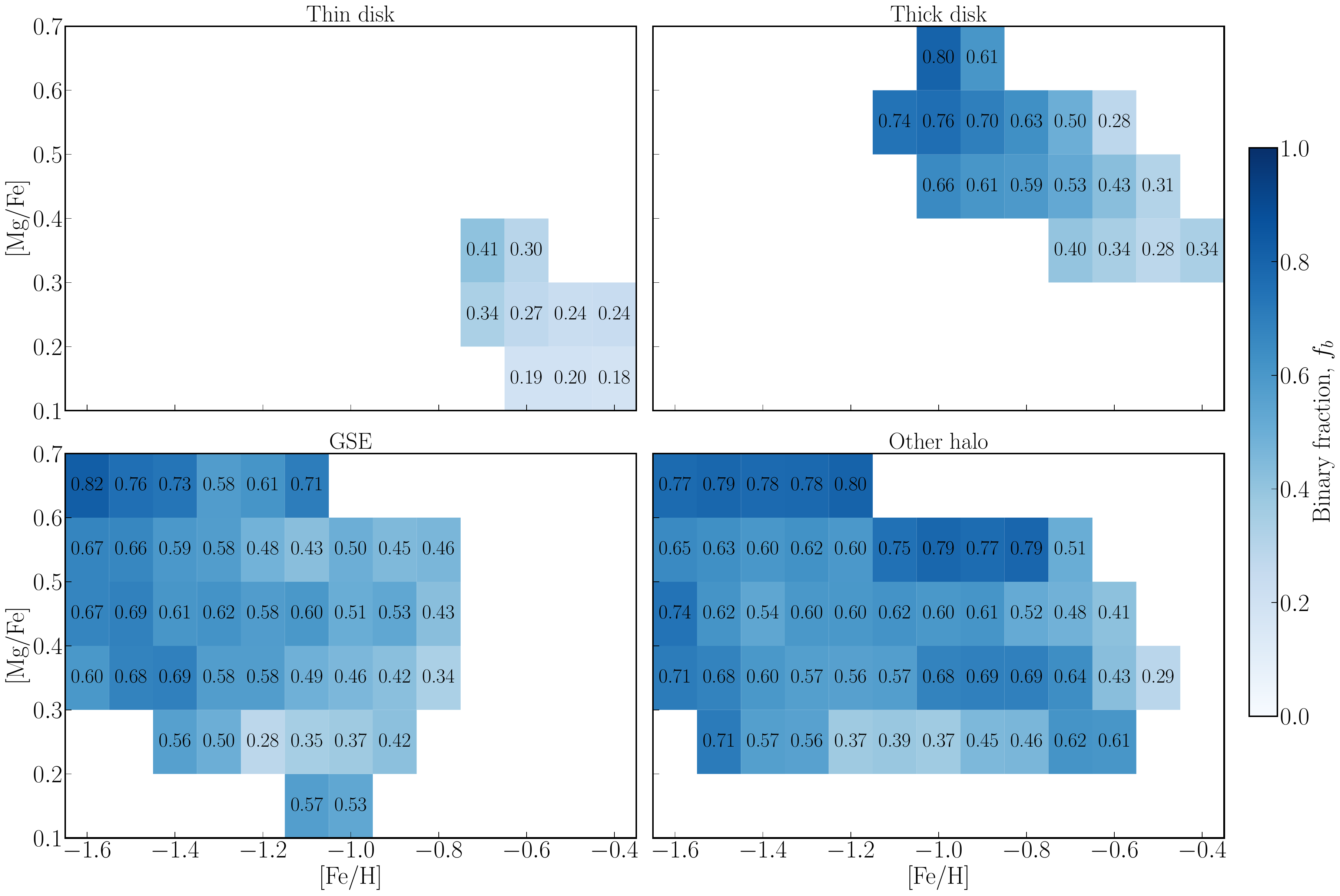}
\caption{
Binary fraction across the [Fe/H]–[Mg/Fe] plane for different Galactic components. The stellar sample is divided into chemically defined thin disk, thick disk, GSE, and other halo populations. Within each component, the binary fraction is inferred independently in chemical-abundance bins. The numbers in each bin denote the median inferred binary fraction, while the background color indicates the corresponding value using a common color scale for all four panels. Empty bins correspond to regions containing too few stars to provide a reliable constraint.
\label{fig:component_fb_maps}}
\end{figure*}

The binary fraction maps over the [Fe/H]--[Mg/Fe] plane are shown in Figure~\ref{fig:component_fb_maps} for different components. For each Galactic component, the subsamples in the [Fe/H]--[Mg/Fe] bins are constructed using the same Monte Carlo resampling procedure adopted in Section~\ref{ssec:mgfe}.

The four components occupy markedly different regions of chemical-abundance space. The thin disk is confined primarily to the metal-rich, low-[Mg/Fe] region and generally exhibits relatively low binary fractions. The thick disk extends toward lower [Fe/H] and higher [Mg/Fe] and shows higher binary fractions than the thin disk in most populated bins. Both GSE and the other-halo component occupy predominantly metal-poor regions and typically exhibit relatively high binary fractions. Within each component, the overall level of $f_b$ generally increases toward decreasing [Fe/H], consistent with the [Fe/H] dependence found for the full sample. The dependence on [Mg/Fe] is less consistent, with an increasing trend present in some [Fe/H] ranges but weak or absent in others, and with no common trend across the different Galactic components.

\begin{figure}
\plotone{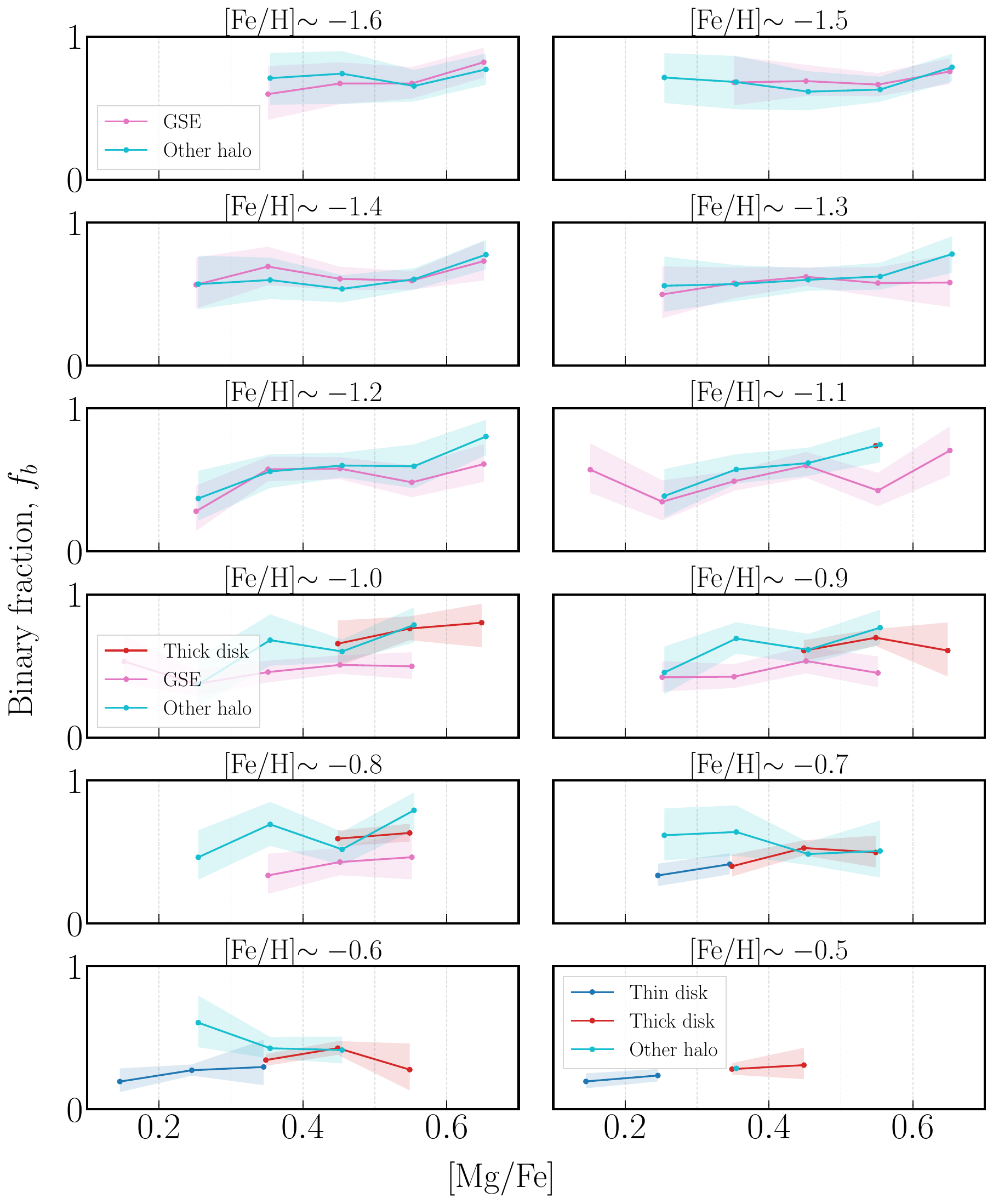}
\caption{
Comparison of the binary fraction among Galactic components as a function of [Mg/Fe] at fixed [Fe/H]. Each panel corresponds to a subsample centered at the indicated [Fe/H], constructed using the Monte Carlo resampling procedure described in Section~\ref{ssec:mgfe}. 
Different colors represent the thin disk, thick disk, GSE, and the other-halo components. Points show the inferred binary fractions, while the shaded regions
indicate their associated uncertainties. A component is shown only where sufficient stars are available for a reliable measurement. The comparison reveals higher binary fractions toward lower [Fe/H] in all components, but no uniform component-dependent offset at fixed [Fe/H] and [Mg/Fe].
\label{fig:component_fb_slices}}
\end{figure}

For a more direct comparison, Figure~\ref{fig:component_fb_slices} presents $f_b$ as a function of [Mg/Fe] for the different components at fixed [Fe/H]. At the metal-rich end, $[\mathrm{Fe}/\mathrm{H}]\simeq-0.5$ to $-0.7$, the thin-disk measurements are restricted to low [Mg/Fe] and generally yield lower binary fractions than the thick-disk and other-halo measurements. However, the components have only limited overlap in [Mg/Fe] in this [Fe/H] range, so the observed differences cannot be attributed uniquely to component membership.

At $-1.2\lesssim[\mathrm{Fe}/\mathrm{H}]\lesssim-0.8$, the thick disk, GSE, and other halo overlap more extensively in abundance space. The other-halo component often exhibits a higher binary fraction than GSE in individual bins, while the thick-disk measurements are generally comparable to those of the other halo at high [Mg/Fe]. Nevertheless, the relative ordering changes with both [Fe/H] and [Mg/Fe], and many of the measurements remain consistent within their uncertainties. There are therefore no component-dependent changes of binary fractions at given when [Fe/H] and [Mg/Fe] are fixed.

At $[\mathrm{Fe}/\mathrm{H}]\lesssim-1.3$, the GSE and other-halo measurements follow one another across most of the available [Mg/Fe] range. Both components exhibit high binary fractions, but neither shows a clear and consistently monotonic dependence on [Mg/Fe]. The bin-to-bin variations become increasingly prominent in this regime because of the smaller numbers of stars available for the resampling analysis.

In summary, Figures~\ref{fig:component_fb_maps} and
\ref{fig:component_fb_slices} indicate that the principal differences among the binary fractions are associated with the distinct [Fe/H] and [Mg/Fe] distributions of the Galactic populations.
The increase of $f_b$ towards lower [Fe/H] remains visible within the individual components, whereas no consistent component ordering is observed when the comparison is restricted to overlapping regions of chemical-abundance space. Any additional dependence on Galactic-component membership therefore appears weaker than the dominant dependence on chemical abundance.

\section{Discussions}
\label{sec:discussion}

The inverse relation between binary fraction and [Fe/H] found in this work is consistent with results from several independent methods and surveys. Multi-epoch radial-velocity studies based on LAMOST and APOGEE have reported higher binary fractions among metal-poor solar-type stars, while analyses combining multiple spectroscopic and eclipsing-binary samples have established a strong anticorrelation between close-binary fraction and metallicity \citep{2018RAA....18...52T,2019ApJ...875...61M,2020MNRAS.499.1607M}. A similar dependence was recovered by \citet{2021ApJ...922..211N} using unresolved photometric binaries, despite the different sensitivity of that method to orbital period and mass ratio. The consistency among these studies suggests that metallicity is an important factor influencing the binary population, particularly for the close-binary systems probed by spectroscopic surveys, although differences in the absolute binary fractions are expected because the surveys probe different ranges of orbital parameters.

Previous measurements of the dependence on $\alpha$ abundances are less consistent. \citet{2018RAA....18...52T} found higher binary fractions among old, high-$\alpha$ stars than among low-$\alpha$ populations. More directly, \citet{2021ApJ...922..211N} reported a positive correlation between binary fraction and $[\alpha/\mathrm{Fe}]$ at fixed [Fe/H] for thin-disk stars, while no significant relation was found for the thick disk. This population-dependent behavior is broadly compatible with our finding that the [Mg/Fe] dependence is most apparent in the relatively metal-rich regime and becomes weak at lower [Fe/H].

By contrast, \citet{2020MNRAS.499.1607M} found that the APOGEE close-binary fraction generally decreases with increasing [Mg/H] abundance at [Fe/H] $\sim 0$. They also reported a complex, non-monotonic dependence of the binary fraction on $\alpha$-element abundance. Their results indicate that the binary fraction decreases toward higher $\alpha$-element abundance over part of the chemical-abundance range, but increases again at very high [$\alpha$/H]. The difference from our result may arise partly from the different stellar populations and chemical ranges sampled: their APOGEE dwarf and subgiant sample is concentrated closer to solar metallicity, whereas our DESI MSTO sample contains substantial thick-disk and halo populations. The two analyses also have different radial-velocity precision, temporal sampling, and sensitivity to orbital period. Consequently, a chemical dependence of the period distribution could produce different observed trends even if the underlying total multiplicity were similar. Residual correlations of [Mg/Fe] with age, birth environment, and Galactic population may also contribute, since matching the metallicity distribution does not control all of these properties.

The physical interpretation of the metallicity dependence is not unique because metallicity affects cooling through competing processes. On the one hand, metal-poor gas cools less efficiently due to fewer metal-line cooling channels\citep[e.g.][]{2026MNRAS.548ag375S,2026MNRAS.548ag300C}, and thus the gas clouds have higher temperature and larger Jeans mass. On the other hand, however, radiation-hydrodynamical simulations by \citet{2019MNRAS.484.2341B} instead show that metal-poor gas can cool more efficiently at high densities due to less dust, which promotes fragmentation on small spatial scales, producing a larger close-binary fraction among low-mass stars, with semi-major axis, $a<10$ au. Importantly, the final separations of such close binaries need not correspond directly to their initial fragmentation scales. Related models suggest that companions formed through disk or small-scale fragmentation can subsequently migrate inward through continued accretion and dynamical interactions, populating the close-binary regime probed by radial-velocity surveys \citep[e.g.,][]{2010ApJ...708.1585K,2019ApJ...875...61M}. In this interpretation, [Fe/H] primarily acts as a proxy for the overall metal and dust content of the star-forming gas. Iron itself is unlikely to be the direct agent controlling binary formation.

The origin of a positive relation with [Mg/Fe] is less straightforward. 
[Mg/Fe] traces the aspects of the chemical-enrichment and formation history of the stellar population. High-[Mg/Fe] stars were enriched predominantly by core-collapse supernovae before Type Ia supernovae contributed large amounts of iron and are therefore generally associated with rapid early star formation.
However, how these differences in formation history translate into binary formation and evolution on the small spatial scales relevant to close binaries remains unclear. The observed $f_b$–[Mg/Fe] relation may therefore reflect differences in the formation histories or other correlated birth-environment properties of the stellar populations, rather than a direct effect of Mg abundance itself. A dependence of binary properties on [Mg/Fe] has also been reported by \citet{2021ApJ...922..211N}, although the physical origin of this dependence remains uncertain.

Several aspects of the present measurement should be considered when interpreting the abundance dependence. First, our analysis focuses on MSTO stars, which occupy a relatively narrow mass range of approximately 0.8-1.0 $\msun$. This reduces the impact of the stellar-mass dependence of binary fraction, but also means that our results cannot be directly generalized to stars of different masses. Since the stellar initial mass function and the mass distribution of observed samples may vary with metallicity and Galactic population, differences in the average stellar mass among different surveys could contribute to variations in the reported binary fractions.

Second, our mock simulations show that the DESI multi-epoch LOSV measurements are primarily sensitive to binaries with orbital periods shorter than $\sim10^4$ days. The inferred binary fractions therefore depend on the assumed orbital-period distributions used to extrapolate from the detectable population to the total binary population. If the orbital-parameter distributions vary with chemical abundance, the observed trends could partly reflect changes in the fraction of systems occupying the period range probed by our measurements. Indeed, \citet{2019MNRAS.482L.139E} found that the metallicity dependence of the wide-binary fraction becomes increasingly pronounced at smaller projected separations ($s\lesssim250$ au), while remaining approximately negligible at larger separations. Together with the non-monotonic metallicity dependence of the wide-binary fraction reported by \citet{2021MNRAS.501.4329H}, these findings suggest that the underlying orbital-period distribution may vary with metallicity. Therefore, while our inferred total binary fractions depend on the adopted orbital-period distribution, the observed LOSV variability primarily constrains the close-binary population accessible to multi-epoch spectroscopy.

We do have also explored models allowing the orbital-period distribution to vary with metallicity. However, the binary fraction and period-distribution parameters exhibit a strong degeneracy, preventing us from obtaining independent constraints on their metallicity dependence with the current observations. We therefore adopt a common lognormal period distribution across all chemical-abundance subsamples in our fiducial analysis. This provides a consistent framework for comparing the inferred binary fractions among different stellar populations, but the resulting binary fractions remain model-dependent. Unfortunately, this is something we have to live with in our current analysis. In the future, with additional observations covering longer time baselines and sensitive to a broader range of orbital periods, it will be more doable and important to constrain the metallicity dependent orbital period distribution and the binary fractions through joint modeling.

Residual variations with Galactic orbital properties may contain information about the assembly and dynamical evolution of the halo. Using Gaia DR3 radial-velocity variability, \citet{2024MNRAS.535..949B} found a moderately higher close-binary fraction among accreted than in-situ halo stars, although the difference was not statistically significant, and it was shown that much of the variation with orbital energy and angular momentum could be associated with differences in metallicity. This is broadly consistent with our result that GSE and other halo populations do not exhibit a clear additional difference in binary fraction once their chemical-abundance distributions are taken into account.

The dependence of binary fraction on different environment has also been investigated in dwarf spheroidal galaxies. \citet{2017AJ....153..254S,2018AJ....156..257S} reported evidence for variations in binary fractions among eight dwarf spheroidal galaxies, suggesting a possible dependence on their formation environments. However, \citet{2023A&A...677A..95A} revisited these measurements and questioned the significance of the reported variations. \citet{2022ApJ...933L..18B} also found indications of a higher close-binary fraction in the Sagittarius dwarf spheroidal galaxy compared to the Milky Way field, based on APOGEE observations. However, these comparisons are complicated by differences in observational sensitivity and stellar populations, particularly because dwarf galaxy studies often rely on evolved giant stars, whose close-binary populations may have been modified by stellar evolution and binary interactions, including Roche-lobe overflow. In contrast, our study uses a large sample of MSTO stars spanning several distinct Galactic components, including both in-situ and accreted populations. The absence of significant differences in their binary fractions at fixed chemical abundance suggests that chemical composition is a major factor associated with the observed variations, with no clear evidence for an additional dependence on Galactic origin within the uncertainties and sensitivity of our analysis.

There may be, nevertheless, additional environmental effects in the inner halo. \citet{2025MNRAS.537.1984A} found that the fraction of carbon-rich stars decreases among more tightly bound field populations to our MW and is particularly low in globular clusters, which they interpreted as evidence for an increasing contribution from disrupted globular clusters toward the inner Galaxy. More directly, \citet{2025MNRAS.541.2008B} showed using Gaia RVS that globular clusters possess substantially lower close-binary fractions than comparable field populations, with little evidence for a metallicity dependence among the clusters themselves. These results suggest that dynamical processes in dense stellar environments can substantially modify the primordial binary population. Such a contribution may help explain some of the residual structure seen in our $L_z$-dependent measurements, including the tentative decrease around $L_z\sim0$, although a more detailed decomposition of the inner halo is required to test this interpretation.

Overall, the metallicity dependence appears to be a robust feature of the close-binary population and is supported by both observations and models. The weaker dependence on [Mg/Fe] is more complex and may combine the effects of detailed chemical composition, formation epoch, and star-formation environment of the binaries. Distinguishing among these possibilities will require simultaneous constraints on both the binary fraction and the orbital-parameter distributions across narrow chemical-abundance bins. Future analyses allowing the period distribution to vary with [Fe/H] and [Mg/Fe] will further help to determine whether the observed chemical dependence reflects an intrinsic variation in the total binary fraction, or whether chemical abundance primarily affects the orbital-period distribution, thereby biasing the inferred binary fraction when a universal period distribution is assumed.

\section{Conclusion}
\label{sec:conclusion}

We have investigated the dependence of the binary star fractions on chemical abundance, Galactic position, and Galactic component/environment (thick/thin disk, halo and GSE) using multi-epoch LOSV measurements from the DESI MWS. Our analysis is based on a volume-limited sample of MSTO stars selected from the MWS \textsc{main-bright} program. Rather than identifying individual binary systems, we infer the binary fraction statistically by comparing the observed velocity-variability distribution with mock populations generated using the actual observing epochs and velocity uncertainties of the DESI data.

Our principal results are summarized as follows.

\begin{enumerate}

\item The binary fraction exhibits a strong dependence on [Fe/H]. Across the two-dimensional [Fe/H]--[Mg/Fe] abundance plane, $f_b$ increases systematically toward lower [Fe/H]. This relation remains evident after controlling for [Mg/Fe] and is therefore the dominant trend detected in our sample.

\item At fixed [Fe/H], the dependence on [Mg/Fe] is substantially weaker. For relatively metal-rich populations with $[\mathrm{Fe}/\mathrm{H}]\gtrsim-1.3$, the binary fraction shows a modest increase with increasing [Mg/Fe]. At lower metallicities, no statistically significant [Mg/Fe] dependence is detected, although the constraints become weaker because fewer stars are available in the metal-poor abundance bins.

\item No clear monotonic dependence of the binary fraction is found with the Galactocentric radius. The full sample exhibits higher binary fractions at larger Galactic heights above the disk, $|z|$, and lower vertical component of the angular momentum, $L_z$, but these trends can largely be associated with the corresponding [Fe/H] gradient. No clear monotonic dependence of the binary fraction is found with either $|z|$ or $L_z$, after controlling for the underlying [Fe/H] distribution to be the same. However, some weak residual variations still remain after controlling for [Fe/H], including a tentative dip around $L_z\sim 0$ and a modest decrease with increasing $|z|$ in some [Fe/H] bins. These residuals broadly follow the variation in [Mg/Fe]. They are more likely associated with the secondary dependence on chemical abundances other than [Fe/H]. 

\item We classify the sample into chemically and kinematically selected thin-disk, thick-disk, GSE, Splash, and other-halo components. The [Fe/H] dependence of the binary fraction remains visible within the individual components, indicating that the global trend is not produced solely by changes in the relative mixture of Galactic populations. When the components are compared in overlapping regions of the [Fe/H]--[Mg/Fe] plane, no prominent differences in their binary fractions are found. Thus chemical abundance is the dominant factor driving the variation in binary fractions among different Galactic components.

\item The inverse relation between binary fraction and [Fe/H] is consistent with most previous observational studies of solar-type stars. It may reflect the influence of metallicity-dependent opacity and cooling on fragmentation in dense protostellar gas or disks, with lower-metallicity environments favoring the formation of relatively close companions. This dependence may have broader implications for stellar populations formed in the early Universe, where metal-poor star-forming environments could have produced different binary populations. Since binary systems play important roles in stellar feedback, chemical enrichment, and transient phenomena, understanding their metallicity dependence is also relevant for interpreting ancient stellar populations and early galactic evolution. The weaker positive dependence on [Mg/Fe] is less straightforward. Since [Mg/Fe] traces enrichment timescale, stellar age, and formation environment, this trend may reflect differences in the birth conditions of high- and low-$\alpha$ populations rather than a direct effect of magnesium on binary formation.

\end{enumerate}

Taken together, our results indicate that [Fe/H] is the primary observable associated with the binary fraction in the DESI MSTO sample. The dependence on [Mg/Fe] is secondary and varies with [Fe/H], while Galactic position and environment contribute little additional predictive power once chemical abundance is controlled for. These findings support a picture in which the binary population is driven mainly by the physical conditions of star formation and subsequently inherited by the different Galactic populations.

The inferred binary fractions remain dependent on the adopted orbital-period, mass-ratio, and eccentricity distribution model. A variation of these distributions with chemical abundance could affect the recovered trends, particularly because multi-epoch LOSV observations are primarily sensitive to relatively close binaries. The forthcoming DESI DR3 will provide a substantially larger number of epochs and longer temporal baselines, enabling improved constraints on the orbital-parameter distributions themselves and allowing tests of whether the observed abundance dependence arises from changes in the overall binary fraction or from variations in orbital properties. Future analyses combining these expanded spectroscopic data sets with abundance-dependent binary population models will provide a more complete picture of binary formation across Galactic history.

The results presented in this paper can be accessed at DOI: \url{https://doi.org/10.5281/zenodo.22742898}.

\begin{acknowledgments}

This work is supported by NSFC (12573022, 12595312, 12273021), the National Key R\&D Program of China (2023YFA1605600, 2023YFA1605601), 111 project (No.\ B20019), and the Office of Science and Technology, Shanghai Municipal Government (grant Nos. 24DX1400100, ZJ2023-ZD-001). We thank the sponsorship from Yangyang Development Fund. The computations of this work are carried on the National Energy Research Scientific Computing Center (NERSC) and the Gravity supercomputer at the Department of Astronomy, Shanghai Jiao Tong University. SK acknowledges support from the Science \& Technology Facilities Council (STFC) grant ST/Y001001/1. 

WW is very grateful for useful discussions with Prof. Fabo Feng, Yang Huang, Haibo Yuan, and Maosheng Xiang. 

This material is based upon work supported by the U.S. Department of Energy (DOE), Office of Science, Office of High-Energy Physics, under Contract No. DE–AC02–05CH11231, and by the National Energy Research Scientific Computing Center, a DOE Office of Science User Facility under the same contract. Additional support for DESI was provided by the U.S. National Science Foundation (NSF), Division of Astronomical Sciences under Contract No. AST-0950945 to the NSF’s National Optical-Infrared Astronomy Research Laboratory; the Science and Technology Facilities Council of the United Kingdom; the Gordon and Betty Moore Foundation; the Heising-Simons Foundation; the French Alternative Energies and Atomic Energy Commission (CEA); the Secretariat of Science, Humanities, Technology and Innovation (SECIHTI) of Mexico; the Ministry of Science, Innovation and Universities of Spain (MICIU/AEI/10.13039/501100011033), and by the DESI Member Institutions: \url{https://www.desi.lbl.gov/collaborating-institutions}. Any opinions, findings, and conclusions or recommendations expressed in this material are those of the author(s) and do not necessarily reflect the views of the U. S. National Science Foundation, the U. S. Department of Energy, or any of the listed funding agencies.

The authors are honored to be permitted to conduct scientific research on I'oligam Du'ag (Kitt Peak), a mountain with particular significance to the Tohono O’odham Nation.

We thank all contributors to this work for their valuable input and support. We thank our DESI publication handler, Alex Krolewski, for his assistance.

This work has made use of data from the European Space Agency (ESA) mission Gaia (\url{https://www.cosmos.esa.int/gaia}), processed by the Gaia Data Processing and Analysis Consortium (DPAC, \url{https://www.cosmos.esa.int/web/gaia/dpac/consortium}). Funding for the DPAC has been provided by national institutions, in particular the institutions participating in the Gaia Multilateral Agreement.

For the purpose of open access, the author has applied a Creative Commons Attribution (CC BY) licence to any Author Accepted Manuscript version arising from this submission. 

\end{acknowledgments}

%
\facilities{Mayall, Gaia}

\software{astropy\citep{2013A&A...558A..33A,2018AJ....156..123A,2022ApJ...935..167A},
emcee\citep{2013PASP..125..306F},
Matplotlib\citep{2007CSE.....9...90H},
NumPy\citep{2020Natur.585..357H},
\textsc{Korg}\footnote{\url{https://github.com/ajwheeler/Korg.jl}} \citep{2023AJ....165...68W,2024AJ....167...83W},
DESI MWS \textsc{rvs} pipeline\footnote{\url{https://github.com/segasai/rvspecfit}} \citep{2019ascl.soft07013K,2023ApJ...947...37C}.
}
          



\bibliography{sample701}{}
\bibliographystyle{aasjournal7}



\end{document}